\documentclass{aa}  
\usepackage{siunitx}
\usepackage{graphicx}
\usepackage{txfonts}
\usepackage{natbib}
\usepackage{lscape}
\usepackage[
        colorlinks=true,
        citecolor=blue,
        linkcolor=blue]{hyperref}
\usepackage{color}

\begin{document}

   \title{Tayler-Spruit dynamo in stably stratified rotating fluids: Application to proto-magnetars}


   \author{P. Barrère\thanks{\email{paul.barrere@cea.fr}}\inst{1} \and J. Guilet\inst{1} \and R. Raynaud\inst{2} \and A. Reboul-Salze\inst{3}}

   \institute{\inst{1}Université Paris-Saclay, Université Paris Cité, CEA, CNRS, AIM, 91191, Gif-sur-Yvette, France\\
   \inst{2}Universit\'e Paris Cit\'e, Universit\'e Paris-Saclay, CNRS, CEA, AIM, F-91191 Gif-sur-Yvette, France\\
   \inst{3}Max Planck Institute for Gravitational Physics (Albert Einstein Institute), D-14476 Potsdam, Germany\\}

   \date{Received ?????; accepted ?????}

 
  \abstract{
  The formation of highly magnetized young neutron stars, called magnetars, is still a strongly debated question. A promising scenario invokes the amplification of the magnetic field by the Tayler-Spruit dynamo in a proto-neutron star (PNS) spun up by fallback. \citet{barrere2023} supports this scenario by demonstrating that this dynamo can generate magnetar-like magnetic fields in stably stratified Boussinesq models of a PNS interior. To further investigate the Tayler-Spruit dynamo, we perform 3D-MHD numerical simulations with the MagIC code varying the ratio between the Brunt-V{\"a}is{\"a}l{\"a} frequency and the rotation rate.
  We first demonstrate that a self-sustained dynamo process can be maintained for a Brunt-V{\"a}is{\"a}l{\"a} frequency about 4 times higher than the angular rotation frequency. The generated magnetic fields and angular momentum transport follow the analytical scaling laws of~\citet{fuller2019}, which confirms our previous results. We also report for the first time the existence of an intermittent Tayler-Spruit dynamo. For a typical PNS Brunt-V{\"a}is{\"a}l{\"a} frequency of \SI{e3}{s^{-1}}, the axisymmetric toroidal and dipolar magnetic fields range between $\SI{1.2e15}{}-\SI{2e16}{G}$ and $\SI{1.4e13}{}-\SI{3e15}{G}$, for rotation periods of $1-\SI{10}{ms}$.
  Thus, our results provide numerical evidence that our scenario can explain the formation of magnetars. As the Tayler-Spruit dynamo is often invoked for the angular momentum transport in stellar radiative zones, our results are also of particular importance in this field and we provide a calibration of the Fuller et al.'s prescription based on our simulations, with a dimensionless normalisation factor of the order of $\SI{e-2}{}$.}

\maketitle
\section{Introduction}
Soft gamma repeaters and anomalous X-ray pulsars are two classes of neutron stars (NSs) that exhibit a wide variety of high-energy emissions from short chaotic bursts during outbursts phases~\citep[e.g.][]{gotz2006,younes2017,coti2018,coti2021b} to giant flares~\citep{evans1980,hurley1999,hurley2005,svinkin2021}, which are the brightest events observed in the Milky Way. These neutron stars are called magnetars because their emissions have been shown to be powered by the dissipation of their ultra-strong magnetic fields~\citep{kouveliotou1994}. Indeed, these emissions show that they rotate with periods of $\SI{2}{}-\SI{12}{s}$ and have stronger rotation braking
than typical NSs~\citep[e.g.][]{rea2012,olausen2014}. If the spin-down is due to the extraction of rotational energy by a magnetic dipole, we can infer that most magnetars exhibit a surface magnetic dipole of $\SI{e14}{}-\SI{e15}{G}$, which are the strongest known in the Universe. Three magnetars, however, display weaker magnetic dipoles of $\SI{6e12}{}-\SI{4e13}{G}$~\citep{rea2010,rea2012,rea2013,rea2014} but absorption lines detected in the X-ray spectra of two of these magnetars suggest the presence of stronger non-dipolar magnetic fields of $\SI{2e14}{}-\SI{2e15}{G}$~\citep{tiengo2013,rodriguez2016}. These `low-field' magnetars therefore demonstrate that an ultra-strong surface magnetic dipole is not necessary for a neutron star to produce magnetar-like emission.

Magnetars are also suspected to be the central engine of extreme events.  In combination with a millisecond rotation period, magnetars in their proto-neutron star (PNS) stage may power magnetorotational supernova (SN) explosions which are more energetic than typical neutrino-driven SNe~\citep[e.g.][]{burrows2007,dessart2008,takiwaki2009,kuroda2020,bugli2020,bugli2021,bugli2023,Obergaulinger2020,Obergaulinger2021,Obergaulinger2022}. The formation of a millisecond magnetar is a popular scenario to explain super-luminous SNe~\citep{woosley2010,kasen2010,dessart2012,inserra2013,nicholl2013} and hypernovae, of which the latter are associated to long gamma-ray bursts ~\citep[GRBs ; ][]{duncan1992,zhang2001,woosley2006,drout2011,nomoto2011,gompertz2017,metzger2011,metzger2018}. In the case of binary NS mergers, the NS remnant may be a magnetar whose magnetic fields power the plateau phase observed in short GRBs afterglows~\citep[e.g.][]{lu2014,gompertz2014,kiuchi2023}.

Recently, the observation of the fast radio burst FRB~200428 was associated to X-ray bursts of the magnetar SGR~1935+2154~\citep{bochenek2020,chime2020,mereghetti2020,zhu2023,tsuzuki2024}, which supports magnetar-powered emission scenarios to explain at least a fraction of FRBs.

In order to better understand these phenomena, it is thus essential to investigate the question of magnetar formation and especially the origin of their ultra-strong magnetic fields. The association of a few magnetars with SN remnants suggests that they are born during core-collapse SNe~\citep{vink2006,martin2014,zhou2019}. The magnetic fields could be amplified during the core-collapse due to the magnetic flux conservation~\citep{ferrario2006,hu2009,schneider2019,shenar2023}. However, the magnetic field of the iron core of the progenitor star is not constrained by observations and it is uncertain whether this scenario can explain the whole magnetar population~\citep{makarenko2021}. A second type of scenario invokes a dynamo action in the newly formed PNS to generate strong large-scale magnetic fields. Three mechanisms have been studied: the convective dynamo~\citep{thompson1993,Raynaud2020,raynaud2022,masada2022,white2022}, the magnetorotational instability (MRI)-driven dynamo~\citep{obergaulinger2009,moesta2014,rembiasz2017,reboul2021a,reboul2022,guilet2022}, and the Tayler-Spruit dynamo~\citep{barrere2022,barrere2023}.

The two former dynamos have been shown to form magnetar-like magnetic fields in the case of fast rotation. In the framework of the millisecond magnetar model, these mechanisms are therefore promising to explain the formation of the central engine extreme explosions. Nevertheless, two uncertainties remain. First, SN remnants associated with magnetars show typical explosion energy of $\sim\SI{e51}{erg}$. This implies that most magnetars were born in standard core-collapse supernovae (CCSNe), which require slower rotation periods of at least $\SI{5}{ms}$~\citep{vink2006}. Second, the rotation is assumed to stem from a fast-rotating progenitor core. It is unclear whether there is a large enough fraction of these progenitors to explain the entire magnetar population.

To address these points, we developed in~\citet{barrere2022} a new formation scenario in which the PNS rotation is determined by the fallback, which is the matter that is initially ejected by the SN explosion before eventually falling back onto the PNS. 3D CCSNe numerical simulations show that the fallback accretion can significantly spin up the PNS surface~\citep{chan2020,stockinger2020,janka2021}. We argued that the differential rotation caused by this spin-up triggers the development of the Tayler-Spruit dynamo. This dynamo mechanism is driven by the combination of the differential rotation and the Tayler instability, which is an instability triggered by perturbations of a purely toroidal magnetic field~\citep{tayler1973,goossens1980a,goossens1981}. Studies in stellar evolution~\citep[such as][]{eggenberger2005,cantiello2014,eggenberger2019a,eggenberger2019b,denhartog2020,griffiths2022} often rely on this mechanism to explain the strong angular momentum transport (AMT) inferred via asteroseismology in sub-giant/red giant stars~\citep[e.g.][]{mosser2012,deheuvels2014,deheuvels2015,gehan2018}. As our model in~\citet{barrere2022}, these works use analytical prescriptions to model the AMT produced by the large-scale magnetic fields in stellar radiative zones. Two distinct analytical models of the Tayler-Spruit dynamo are used: the original model of~\citet{spruit2002} and the revised one of~\citet{fuller2019}, which tackles previous criticism of the original model. The prescriptions, however, can not take into account the strong non-linearity behind the dynamo mechanism, which impels numerical investigations of its 3D complex dynamics to better characterize its effects in both astrophysical contexts.

\citet{petitdemange2023,petitdemange2024,daniel2023} performed 3D direct numerical simulations of dynamo action in a stably-stratified Couette flow in the context of stellar radiative layers and so with a differential rotation in which the inner core rotates faster than the outer layer. They argued that the dynamo was driven by the Tayler instability and found scaling laws in agreement with the prescriptions of ~\citet{spruit2002}. In our numerical study~\citet{barrere2023}, the setup is different with an outer sphere rotating faster than the core, which is relevant to our magnetar formation scenario. We demonstrated the existence of more complex dynamics with two Tayler-Spruit dynamo branches, which have distinct magnetic field strengths and geometries: the weaker branch shows a hemispherical field while the strongest one displays a dipolar symmetry, i.e. the magnetic field is equatorially symmetric. Furthermore, the former follows the analytical scaling of~\citet{spruit2002}, while the latter is in agreement with the predictions of~\citet{fuller2019}. Lastly, the dipolar dynamo could reach axisymmetric toroidal and dipole magnetic fields up to $\sim \SI{2e15}{G}$ and $\sim \SI{3e14}{G}$, respectively. Although such intensities seem relevant to form magnetars, these models considered only a fixed ratio of the Brunt-V{\"a}is{\"a}l{\"a} to the surface angular frequency $N/\Omega_o=0.1$, whereas it is expected to cover the range $N/\Omega_o\in[0.1,10]$ in real PNSs.

Therefore, this article aims at investigating the impact of $N/\Omega_o$ on the dipolar Tayler-Spruit dynamo discovered in~\citet{barrere2023}. The study of the hemispherical dynamo will lead to another paper more focused on the complex physics behind the Tayler-Spruit dynamos. In the following, Sect.~\ref{sec:setup} describes the governing equations and the numerical setup of our simulations. We present the results in Sect.~\ref{sec:results}, which will be applied to the question of magnetar formation in Sect.~\ref{sec:magnetars}. Finally, we discuss the results and conclude in Sect.~\ref{sec:discussion} and Sect.~\ref{sec:conclusions}, respectively. 


\begin{figure*}
    \centering
    \includegraphics[width=0.65\textwidth]{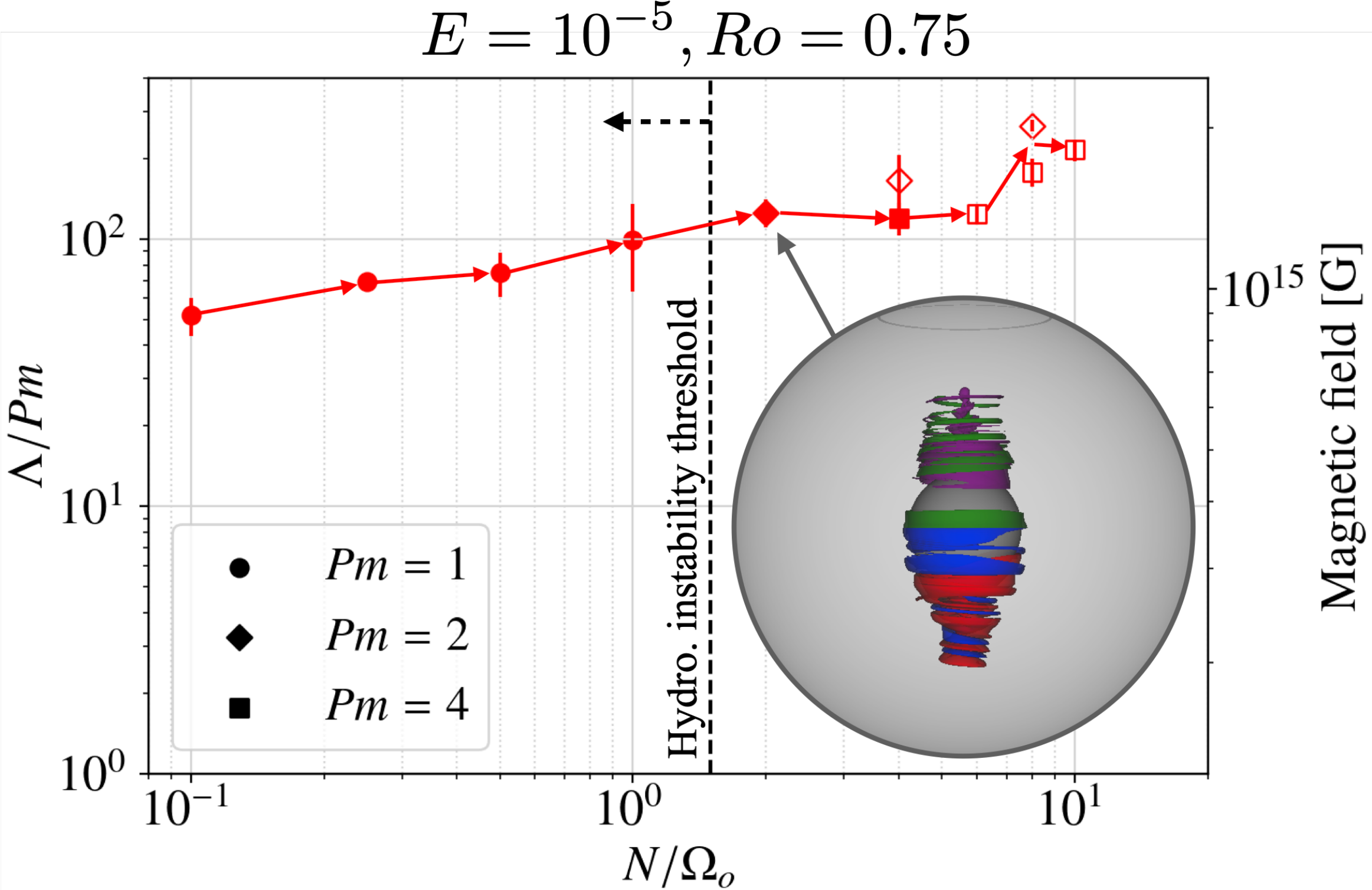}
    \caption{Viscous Elsasser number (and root mean square magnetic field) as a function of the ratio of the Brunt-V\"{a}is\"{a}l\"{a} frequency to the rotation rate at the outer sphere. Filled and empty markers represent self-sustained and transient dynamos, respectively. The black dashed vertical line and arrow indicate the zone in which the fluid is hydrodynamically unstable. The inset represents a 3D plot of the radial velocity (violet and green isosurfaces are the positive and negative values, respectively) and the radial magnetic field (red and blue isosurfaces are the positive and negative values, respectively) in a run at $Pm=2$ and $N/\Omega_o=2$. The grey arrow points to the run location in the diagram.}
    \label{fig:bifurcation}
\end{figure*}

\section{Numerical setup} \label{sec:setup}

\subsection{Governing equations}\label{sec:equations}
As in~\citet{barrere2023}, we model the PNS interior as a stably stratified and electrically conducting fluid. We also adopt the Boussinesq approximation and consider a fluid with a constant density $\rho=\SI{3.8e14}{g.cm^{-3}}$, which corresponds to a PNS with a radius of $r_o=\SI{12}{km}$ and a mass of $M=\SI{1.4}{M_{\odot}}$. The fluid evolves in a spherical Taylor-Couette configuration, i.e. between two concentric spheres of radius~$r_i=\SI{3}{km}$ and $r_o$ which rotates with the respective rates $\Omega_i=2\pi\times25\,{\rm rad\,s^{-1}}$ and $\Omega_o=2\pi\times100\,{\rm rad\,s^{-1}}$. In the reference frame rotating with the surface at the angular velocity $\mathbf{\Omega}_o= \Omega_o \mathbf{e}_z$, the Boussinesq MHD equations read
\begin{align} \label{eq:1}
    \mathbf{\nabla \cdot v} &=0\,,\\ \label{eq:2}
        D_t\mathbf{v} &= -\frac{1}{\rho}{\nabla p'} -2\Omega_o \mathbf{e}_z\times \mathbf{v}
        -N^2\Theta\mathbf{e}_r
         + \frac{1}{4 \pi \rho} (\nabla\times \mathbf{B})\times \mathbf{B} + \nu\Delta\mathbf{v}\,, 
    \\ \label{eq:3}
    D_t\Theta &=\kappa\Delta \Theta\,,\\ \label{eq:4}
    \partial_t\mathbf{B} &=\nabla\times(\mathbf{u}\times\mathbf{B})+\eta\Delta \mathbf{B}\,,\\ \label{eq:5}
    \nabla\cdot\mathbf{B} &=0\,, 
\end{align}
where $\mathbf{B}$ is the magnetic field, $\mathbf{v}$ is the velocity field, $p'$ is the non-hydrostatic pressure, $\rho$ is the mean density of the PNS, $g=g_o r/r_o$ is the gravitation field, and $\alpha\equiv \rho^{-1}(\partial_{T}\rho)_{p}$ is the thermal expansion coefficient. $\mathbf{e}_z$ and $\mathbf{e}_r$ are the unit vectors of the axial and the spherical radial directions, respectively. $\theta$ is the buoyancy variable defined by
\begin{equation}\label{eq:buoyancy}
    \Theta\equiv-\frac{g}{N^2}\frac{\rho'}{\rho}\,,
\end{equation}
where $\rho'$ is the density perturbation due to the combined effect of the electron fraction and entropy perturbations and 
\begin{equation}\label{eq:brunt-vaisala}
    N\equiv \sqrt{-\frac{g}{\rho}\left(\left.\frac{\partial\rho}{\partial S}\right|_{P,Y_e}\frac{dS}{dr}+\left.\frac{\partial \rho}{\partial Y_e}\right|_{P,S}\frac{dY_e}{dr}\right)}
    \,,
\end{equation}
is the Brunt-V\"{a}is\"{a}l\"{a} frequency with the electron fraction $Y_e$, and the entropy $S$, respectively.

In the above equations, we assume that the magnetic diffusivity $\eta$, the kinematic viscosity $\nu$, and the ``thermal'' diffusivity $\kappa$\ are constant. We also assume that the thermal and lepton number diffusivities are equal, which allows us to describe the buoyancy associated with both entropy and lepton number gradients with the use of a single buoyancy variable~$\theta$ \citep{guilet2015b}.

Apart from the magnetic diffusivity which relates to the electrical conductivity of electrons, the physical interpretation of the other transport coefficients can lead to different estimates, depending on whether neutrinos are considered or not to be the main source of diffusive processes~\citep[see Sect. 1.3 of the supplementary materials in][]{barrere2023}.

Finally, we apply no-slip, electrically insulating, and fixed buoyancy variable boundary conditions on both shells. 

\subsection{Numerical methods}\label{sec:numerical}
We use the open source pseudo-spectral code MagIC\footnote{Commit 2266201a5 on \url{https://github.com/magic-sph/magic}} \citep{wicht2002,gastine2012,schaeffer2013} to integrate Eqs.~\eqref{eq:1}--\eqref{eq:5} in 3D spherical geometry. To satisfy the solenoidal conditions~\eqref{eq:1} and \eqref{eq:5}, the velocity and magnetic fields are decomposed in poloidal and toroidal components (Mie representation),
\begin{align}
    \rho\mathbf{u}&=\nabla\times(\nabla\times W\mathbf{e}_r)+\nabla\times Z\mathbf{e}_r\,,\\
    \mathbf{B}&=\nabla\times(\nabla\times b\mathbf{e}_r)+\nabla\times a_j\mathbf{e}_r\,,
\end{align}
where $W$ and $Z$ ($b$ and $a_j$) are the poloidal and toroidal potentials for the velocity (magnetic) field. The whole system of equations is then solved in spherical coordinates by expanding the scalar potentials in Chebyshev polynomials in the radial direction, and spherical harmonic functions in the angular directions. The time-stepping scheme used is the implicit/explicit Runge-Kutta BPR353~\citep{boscarino2013}. We refer the reader to the MagIC online documentation\footnote{\url{https://magic-sph.github.io}} for an exhaustive presentation of the numerical techniques.

\subsection{Input parameters}\label{sec:input_param}
The resistivity is controlled by the magnetic Prandtl number $Pm\equiv\nu/\eta$. Though its realistic value in PNSs~\citep[$Pm\sim 10^{11}$,][]{barrere2023} can not be reached by numerical simulations, we stay in the regime $Pm\geqslant 1$ as we impose $Pm\in[1,4]$. We keep fixed the other dimensionless control parameters: the shell aspect ratio $\chi\equiv r_i/r_o=0.25$ and width $d\equiv r_o-r_i$, the Ekman number $E\equiv\nu/(d^2\Omega_o)=10^{-5}$, 
the thermal Prandtl numbers $Pr\equiv\nu/\kappa=0.1$, 
and the Rossby number $Ro\equiv 1-\Omega_i/\Omega_o = 0.75$, which controls the imposed differential rotation.

The imposed stable stratification is characterized by the Brunt-V\"{a}is\"{a}l\"{a} frequency $N$ (Eq.~\eqref{eq:brunt-vaisala}). In our parameter study, the ratio $N/\Omega_o$ is varied between $0.1$ and $10$ and so covers the PNS regime. In practice, this ratio is related to the Rayleigh number $Ra\equiv -(N/\Omega)^2Pr/E^2$, which is negative in the regime of stable stratification.

The resolution is fixed at $(n_r,n_{\theta},n_{\phi})=(257,256,512)$ for all the runs. A few simulations were rerun with a higher resolution of $(n_r,n_{\theta},n_{\phi})=(481,512,1024)$ but showed no significant change compared to runs with the usual resolution (see Appendix).

The simulations are initialized either from a nearby saturated state or a strong ($B_{\phi}=\SI{3.4e14}{G}$) toroidal axisymmetric field with a dipolar equatorial symmetry, i.e. equatorially symmetric\footnote{For the choice of these definitions, see \cite{gubbins1993}.} with $l=2,m=0$. We define a turbulent resistive time $\bar{\tau}_\eta = \left(\pi r_o/\bar{\ell}\right)^2/\eta\sim0.2 d^2/\eta$, where $\bar{\ell}=10$ is the typical value of the average harmonic degree of the time-averaged magnetic energy spectrum. In the following, we will term a solution `transient' when a steady state is sustained for a time interval $\Delta t > 0.3 \bar{\tau}_\eta$, and `stable' for $\Delta t \geqslant \bar{\tau}_\eta$.

We start with the run named `Ro0.75s' from~\citet{barrere2023}, where the stratification is $N/\Omega_o=0.1$. The saturated state of this dynamo is used to initialise the next simulation with a stronger stratification. The whole set of simulations is initiated similarly using the nearby saturated state of a less stratified run. With this procedure, $N/\Omega_o$ is increased gradually in order to study the evolution of the dynamo branch.

\subsection{Output parameters}\label{sec:output_param}

We first characterize our models by computing the time average of the kinetic and magnetic energy densities (after filtering out any initial transient). The latter is expressed in terms of the viscous Elsasser number $\Lambda_{\nu}\equiv\Lambda/Pm= B_{\rm rms}^2/(4\pi\rho\nu\Omega_o)$ and used to compute different root-mean-square (RMS) estimates of the magnetic field. In addition to the total field, we distinguish the poloidal and toroidal fields based on the Mie representation (Sect.~\ref{sec:numerical}), while the dipole field refers to the $l=1$ poloidal component.
\begin{figure*}
    \centering
    \includegraphics[width=0.49\textwidth]{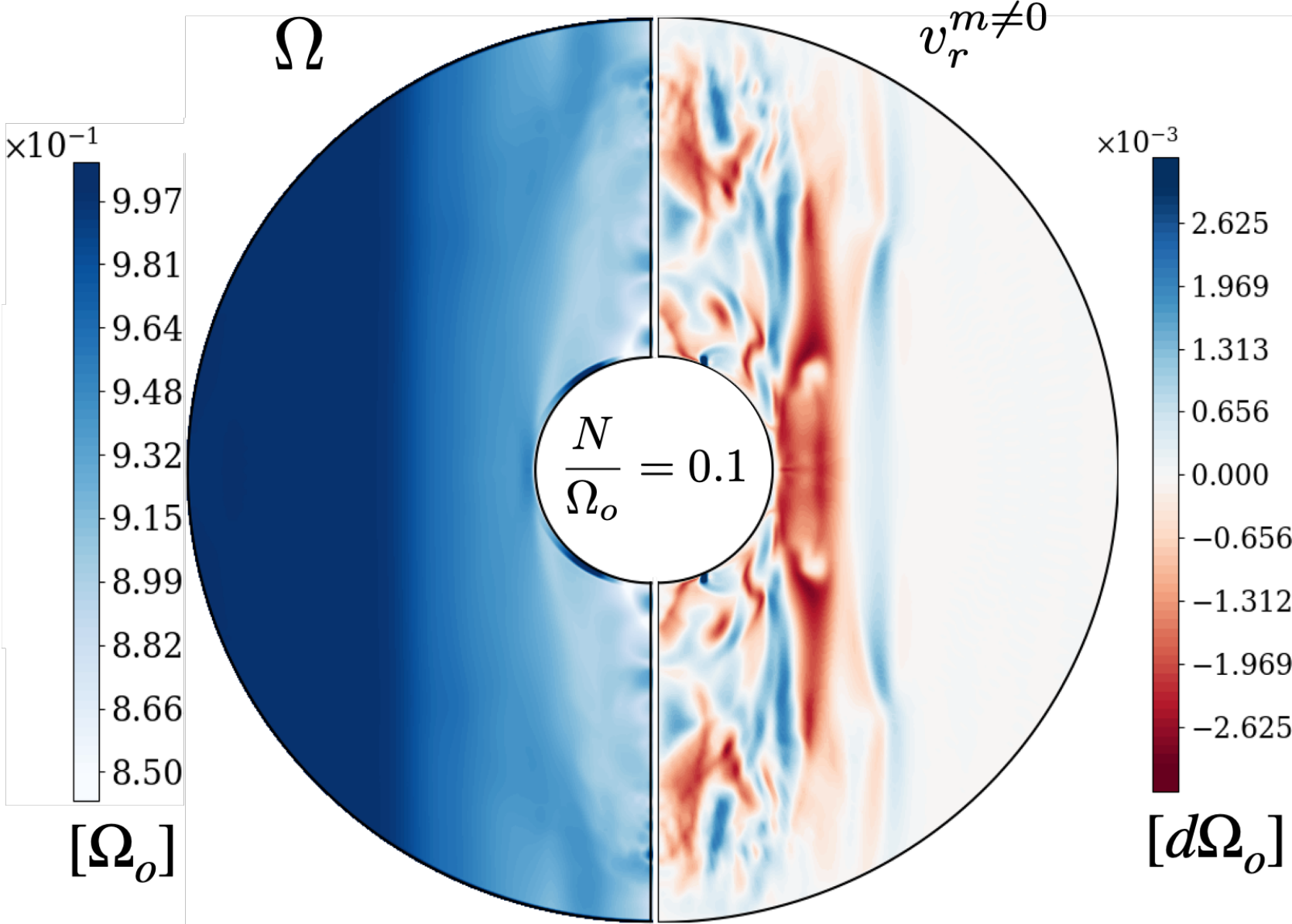}
    \includegraphics[width=0.49\textwidth]{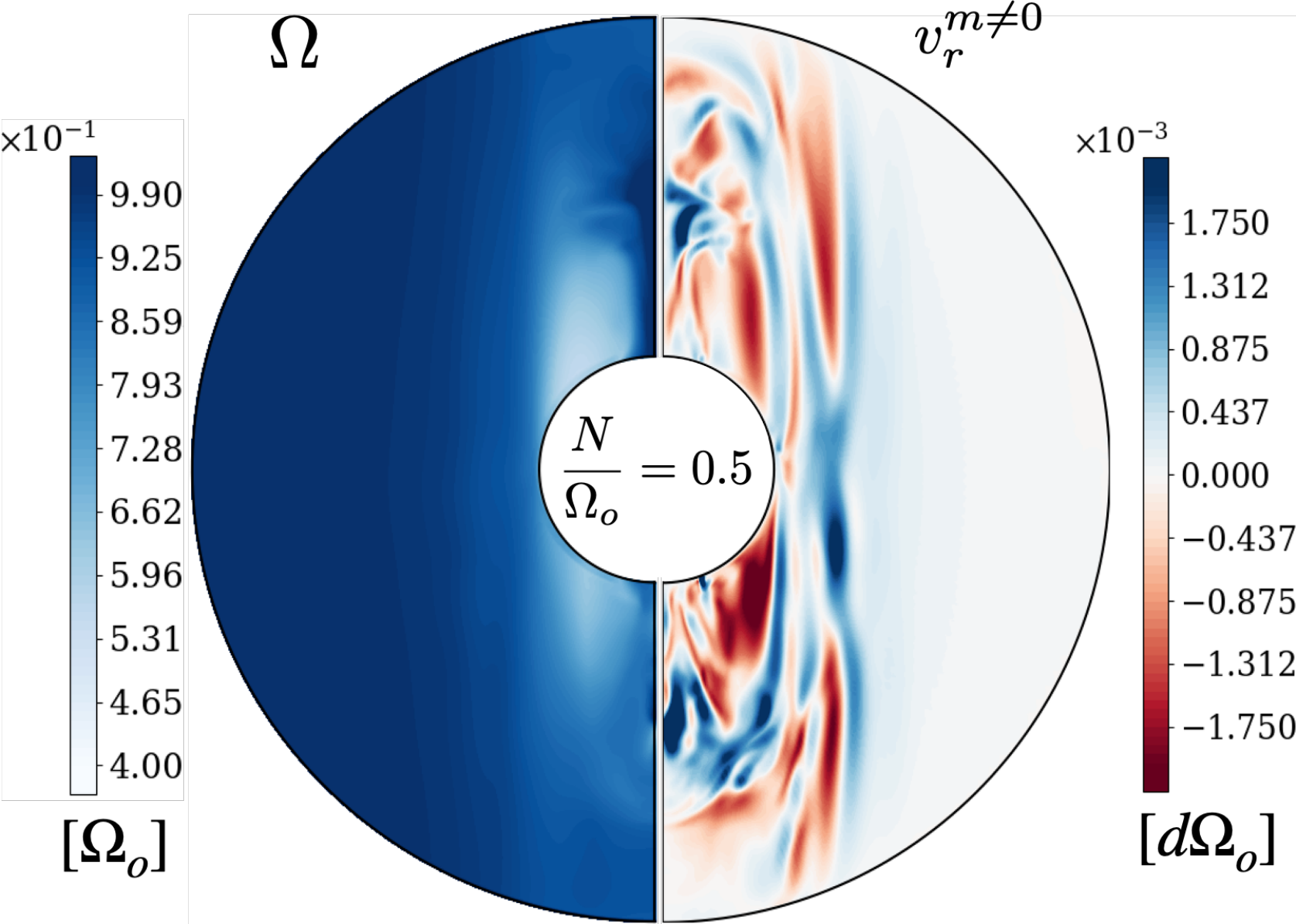}\vspace{0.75cm}
    \includegraphics[width=0.49\textwidth]{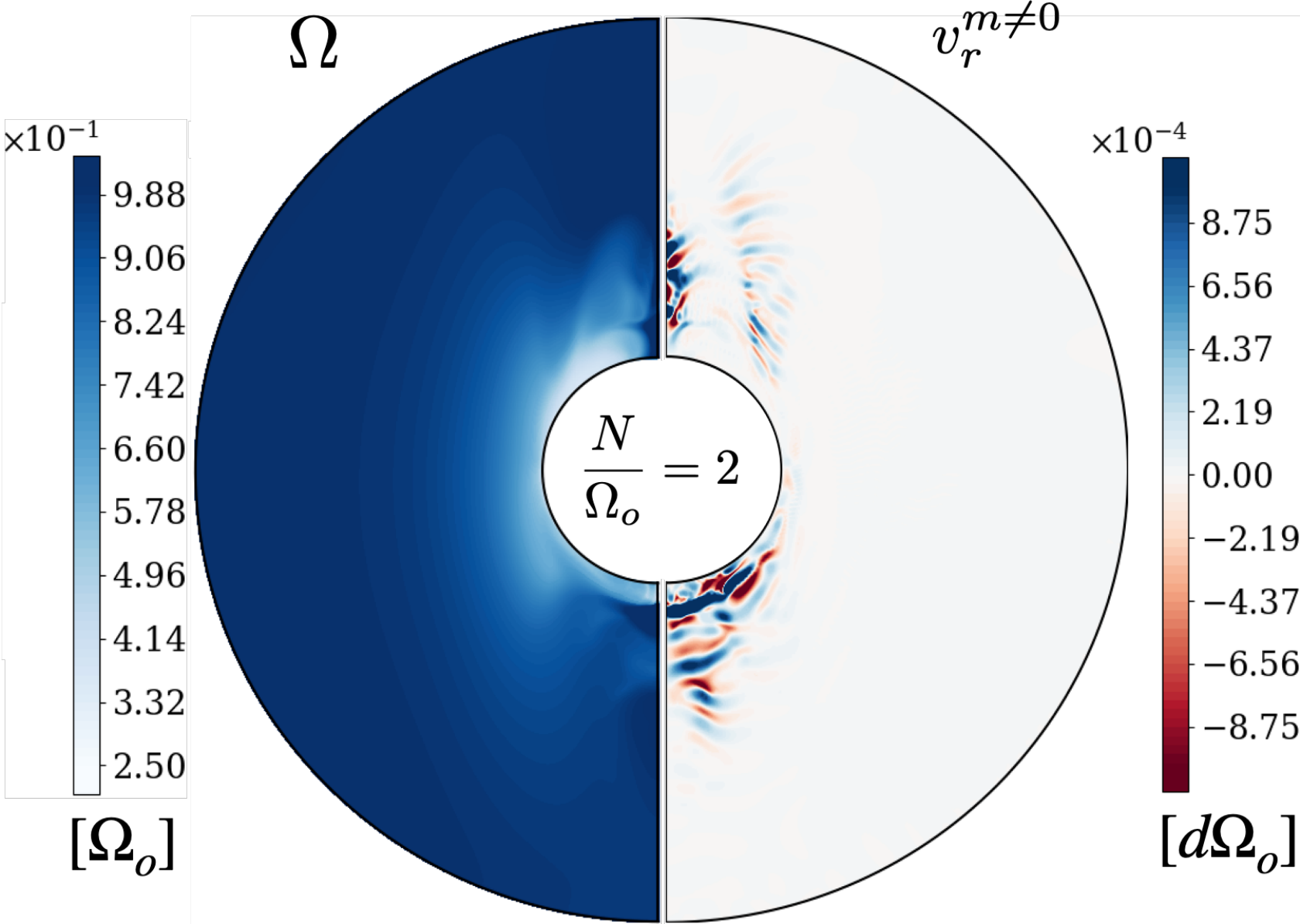}
    \includegraphics[width=0.49\textwidth]{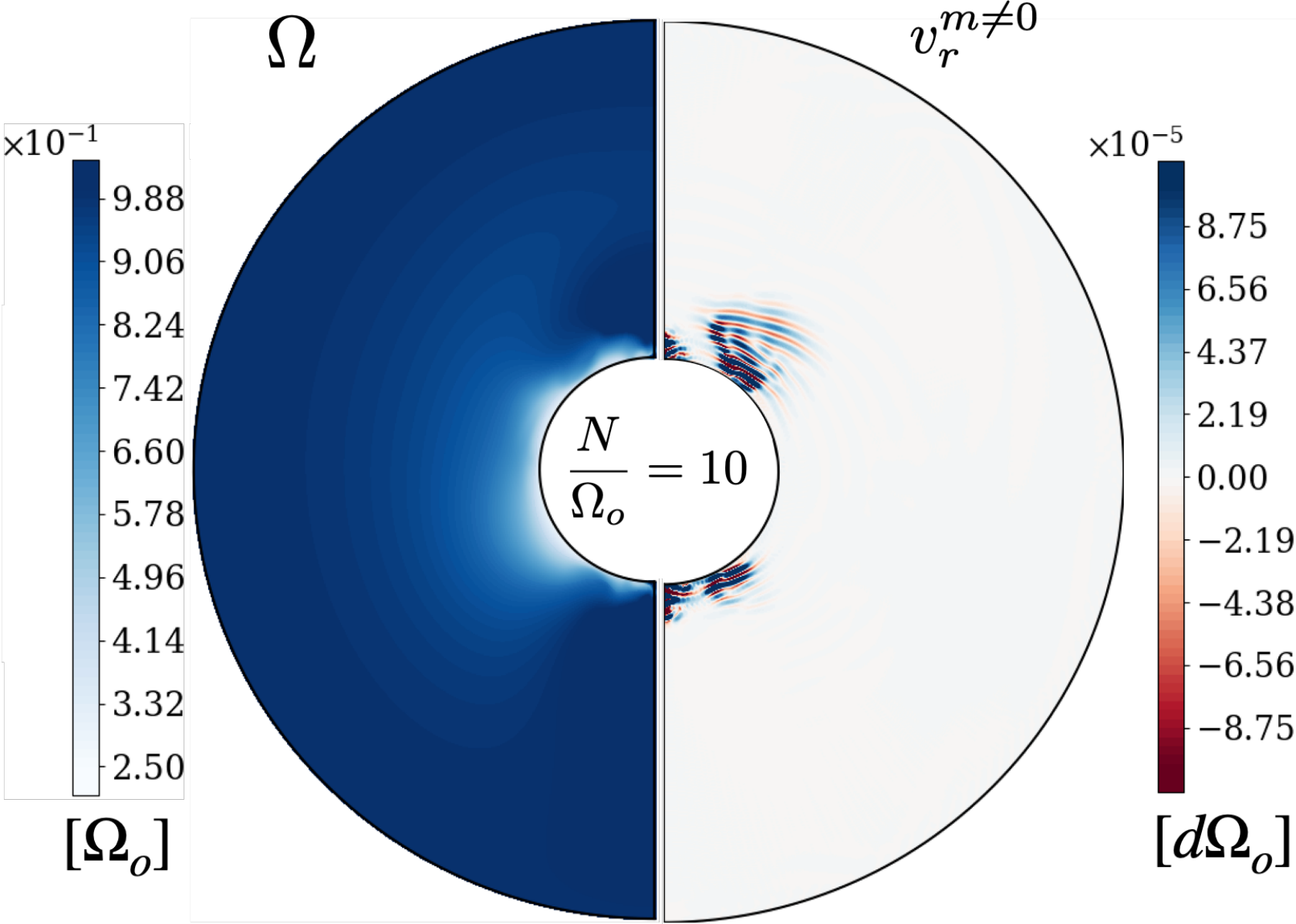}
    \caption{Meridional slices of the angular frequency and the non-axisymmetric radial velocity (left and right slices respectively) for different values of $N/\Omega_o$. $\Omega$ and $v_r^{m\neq 0}$ are scaled  by $\Omega_o$ and $d\Omega_o$, respectively.}
    \label{fig:OmgVrp}
\end{figure*}

\section{Results}\label{sec:results}
The following sections gather the different results we obtain from the set of numerical simulations listed in appendix~\ref{app:models}. We first describe the global dynamics of the dipolar Tayler-Spruit dynamo in the parameter space in Sect.~\ref{sec:subcrit}. Then, we analyse the influence of stratification on the modes of Tayler instability and on the generated axisymmetric magnetic fields in their saturated state in Sect.~\ref{sec:TI} and Sect.~\ref{sec:scalings}, respectively. We also present the angular momentum transport by both Reynolds and Maxwell stresses due to the dynamo and compare the efficiencies of mixing and angular momentum transport in Sect.~\ref{sec:AMT}. Finally, we examine a new intermittent behaviour of the Tayler-Spruit dynamo at $N/\Omega_o\geq 2$, which is observed for the first time (see Sect.~\ref{sec:intermittency}).

\subsection{Subcritical dynamo sustained at PNS-like stratifications}\label{sec:subcrit}
Fig.~\ref{fig:bifurcation} shows that a self-sustained Tayler-Spruit dynamo can be maintained up to $N/\Omega_o=1$ for $Pm=1$. For stronger stratifications, we have to increase $Pm$ (i.e. decrease the resistivity) to maintain the dynamo. For $Pm=4$, the stationary state is self-sustained up to $N/\Omega_o=4$ and we obtained transient states up to $N/\Omega_o=10$. The self-sustained dynamo is therefore present above the threshold for the fluid to be hydrodynamically stable at $N/\Omega_o\sim 1.5$. This confirms the subcritical nature of the Tayler-Spruit dynamo, which was already observed in previous studies~\citep{petitdemange2023,barrere2023}. We did not simulate fluids at greater $Pm$ values for reasons of numerical costs. Given the trend with $Pm$ observed in our simulations as well as theoretical expectations on the Tayler instability threshold, we would expect the Tayler-Spruit dynamo to exist at still higher values of $N/\Omega$ for the higher values of $Pm$ relevant to a PNS.

\subsection{Impact on the differential rotation}\label{sec:diffrot}
\begin{figure*}
    \centering
    \includegraphics[width=0.49\textwidth]{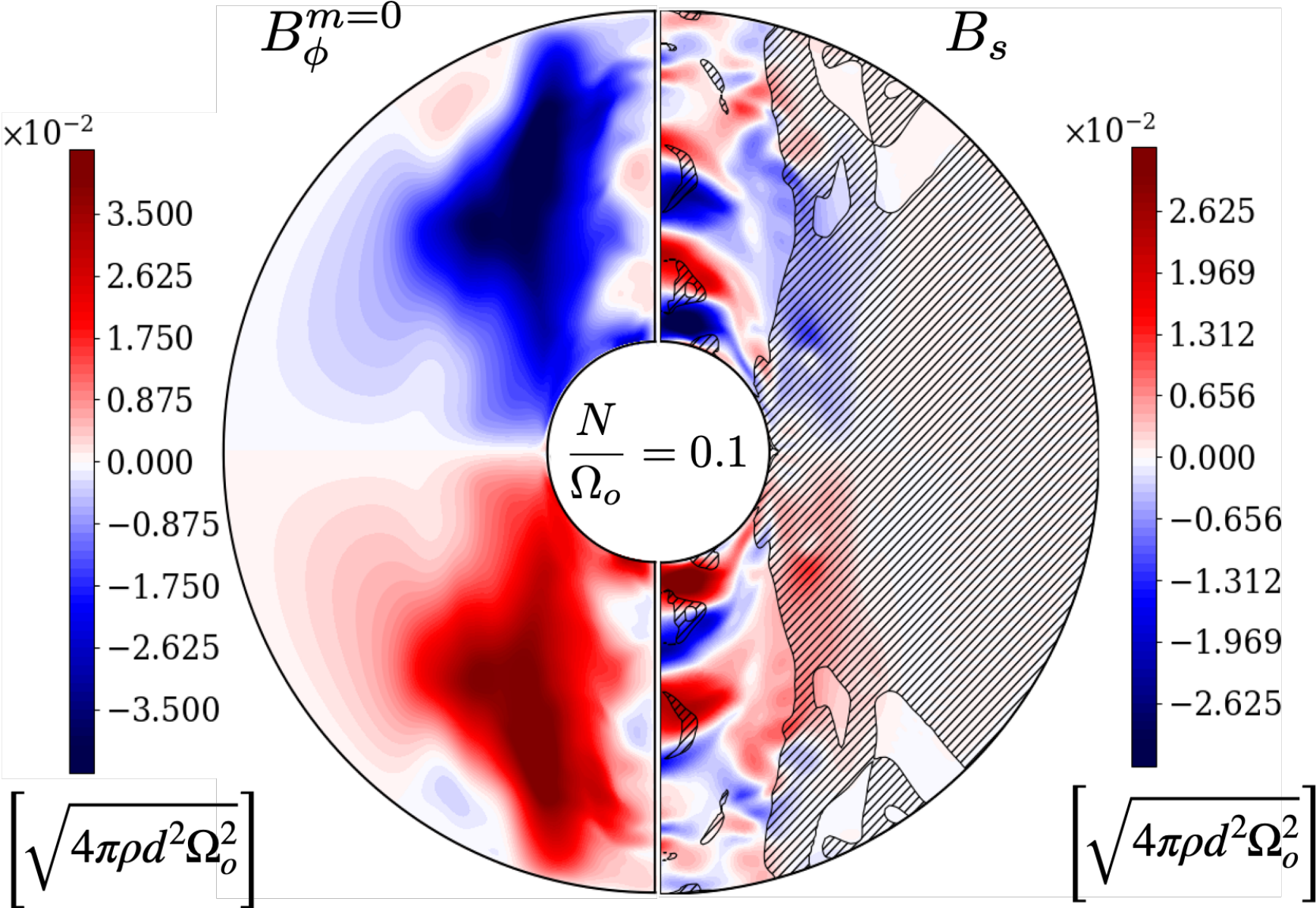}
    \includegraphics[width=0.49\textwidth]{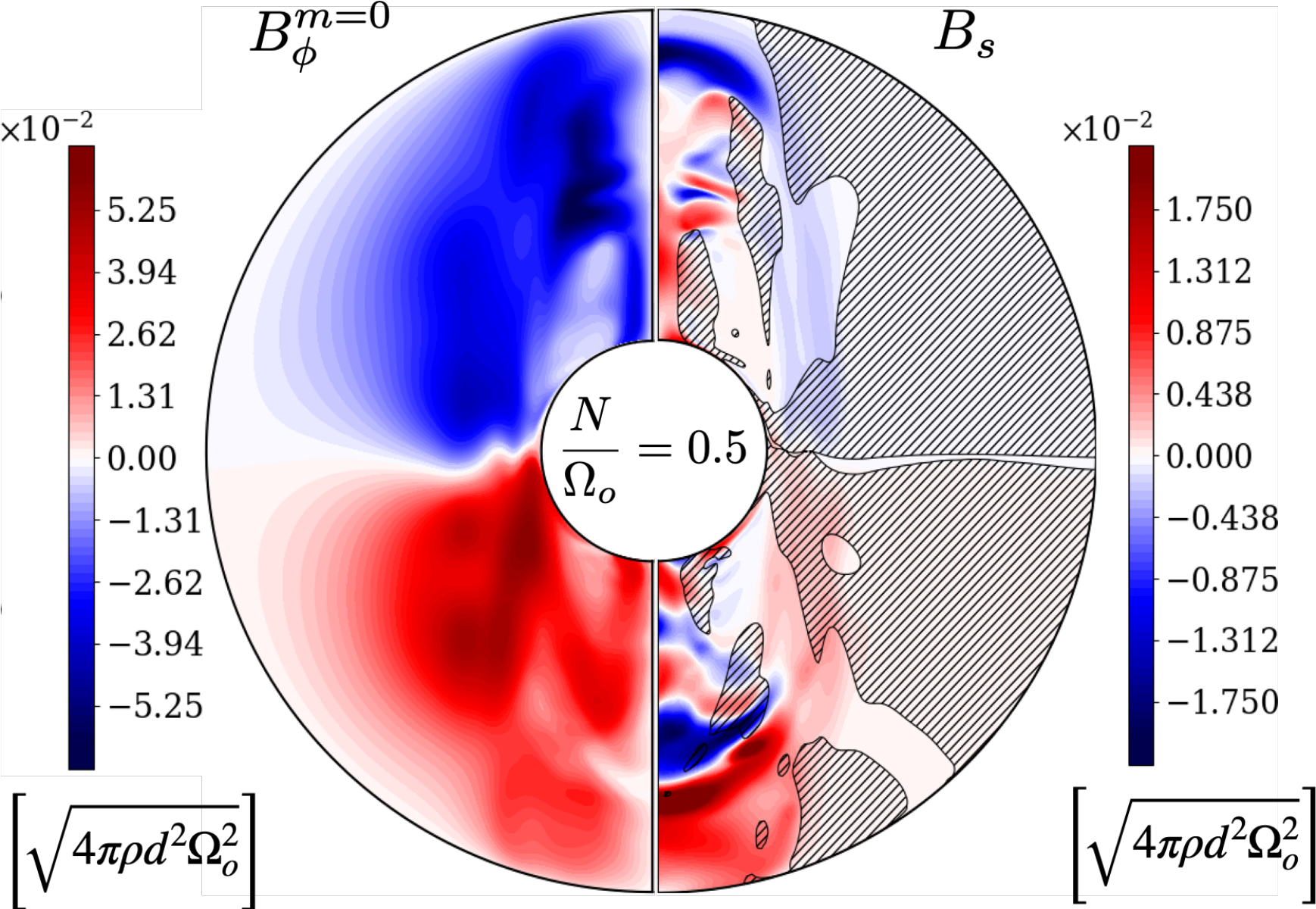}\vspace{0.75cm}
    \includegraphics[width=0.49\textwidth]{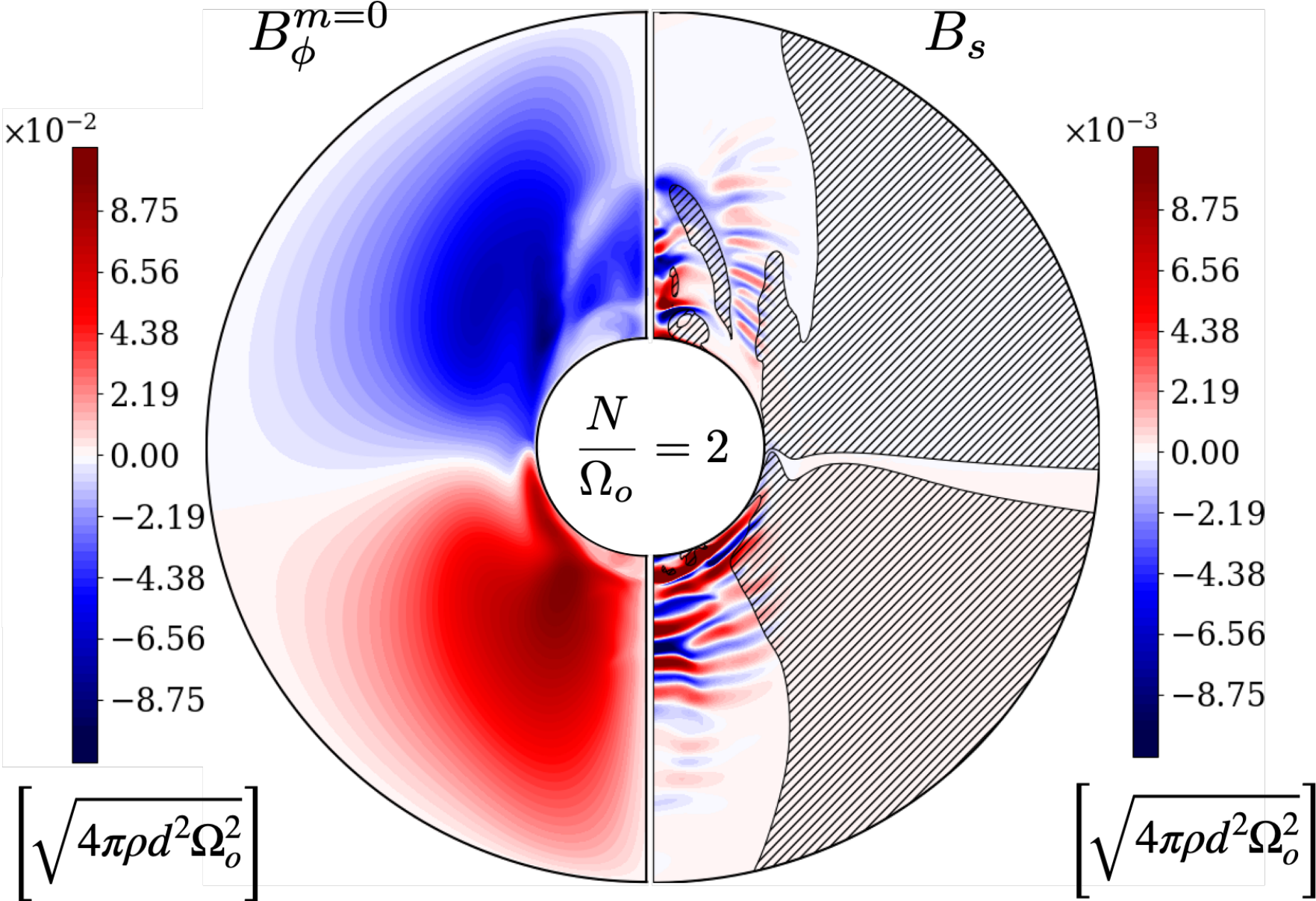}
    \includegraphics[width=0.49\textwidth]{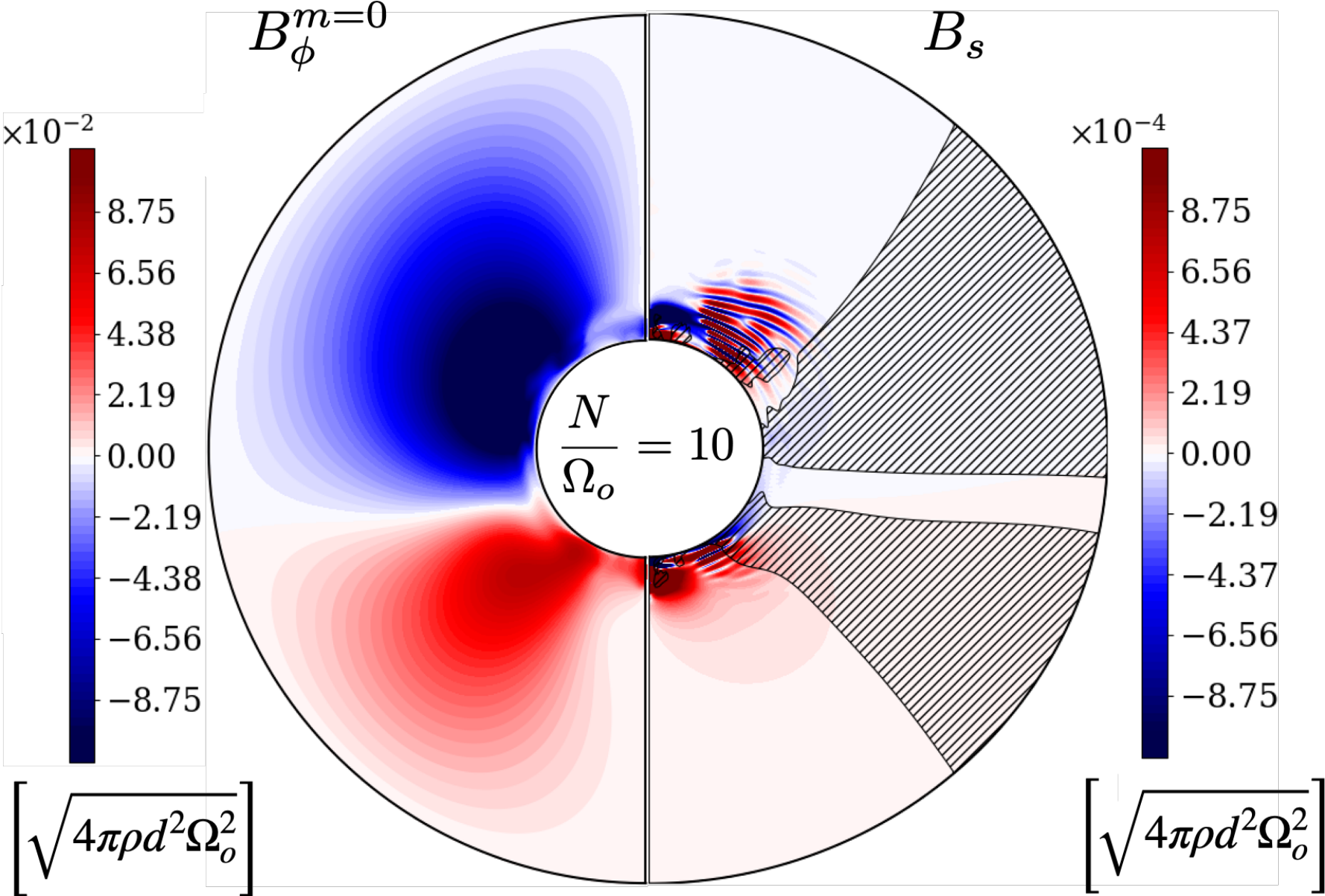}
    \caption{Meridional slices of the axisymmetric azimuthal and the $s=r\sin\theta$-component of the magnetic field (respective left and right slices) for increasing values of $N/\Omega_o$. The magnetic field is scaled by $\sqrt{4\pi\rho d^2\Omega_o^2}$. The hatched regions represent Tayler-stable zones defined by the geometrical criterion of~\citet{goossens1980a} (see Supplementary Materials in~\citet{barrere2023}).}
    \label{fig:BphiBs}
\end{figure*}

The meridional slices of the angular rotation frequency $\Omega$ illustrate the impact of stable stratification on the rotation profile: 
we see that the shear concentrates closer to the inner sphere and increases with $N/\Omega_o$.
At the same time, the rotation profile smoothly transits from a quasi-cylindrical to a spherical geometry, which is an effect already observed in stably stratified flows. 
Analytical and numerical studies of these flows~\citep[e.g.][]{Barcilon1967a,Barcilon1967c,Barcilon1967b,gaurat2015,phillidet2020} indicate that this transition is controlled by the dimensionless parameter $Q\equiv Pr(N/\Omega_o)^2$, which varies between $\SI{e-3}{}$ and $10$ in our set of runs. The change in the flow geometry is therefore explained by a transition from a case where neither the rotation nor the buoyancy dominate ($E^{2/3}<Q<1$) to a buoyancy-dominated flow ($Q\gg 1$).

\subsection{Impact on the Tayler modes}\label{sec:TI}

As seen in Fig.~\ref{fig:BphiBs}, the unstable magnetic modes are located close to the poles where the latitudinal gradient of $B_{\phi}$ is positive, which is a first indication of the presence of Tayler modes. To confirm this statement, we use the geometrical criterion of~\citet{goossens1980a} for the stability of $m=1$-modes,
\begin{equation}
    B_{\phi}^2\left(1-2\cos^2\theta\right) - \sin\theta\cos\theta\frac{\partial B_{\phi}^2}{\partial \theta} > 0\,.
\end{equation}
The stability regions displayed by the hatched zones in Fig.~\ref{fig:BphiBs} match very well regions where the unstable modes are absent. This confirms that the Tayler instability is clearly identifiable, no matter the values of $N/\Omega_o$. 

\begin{figure}
    \centering
    \includegraphics[width=\columnwidth]{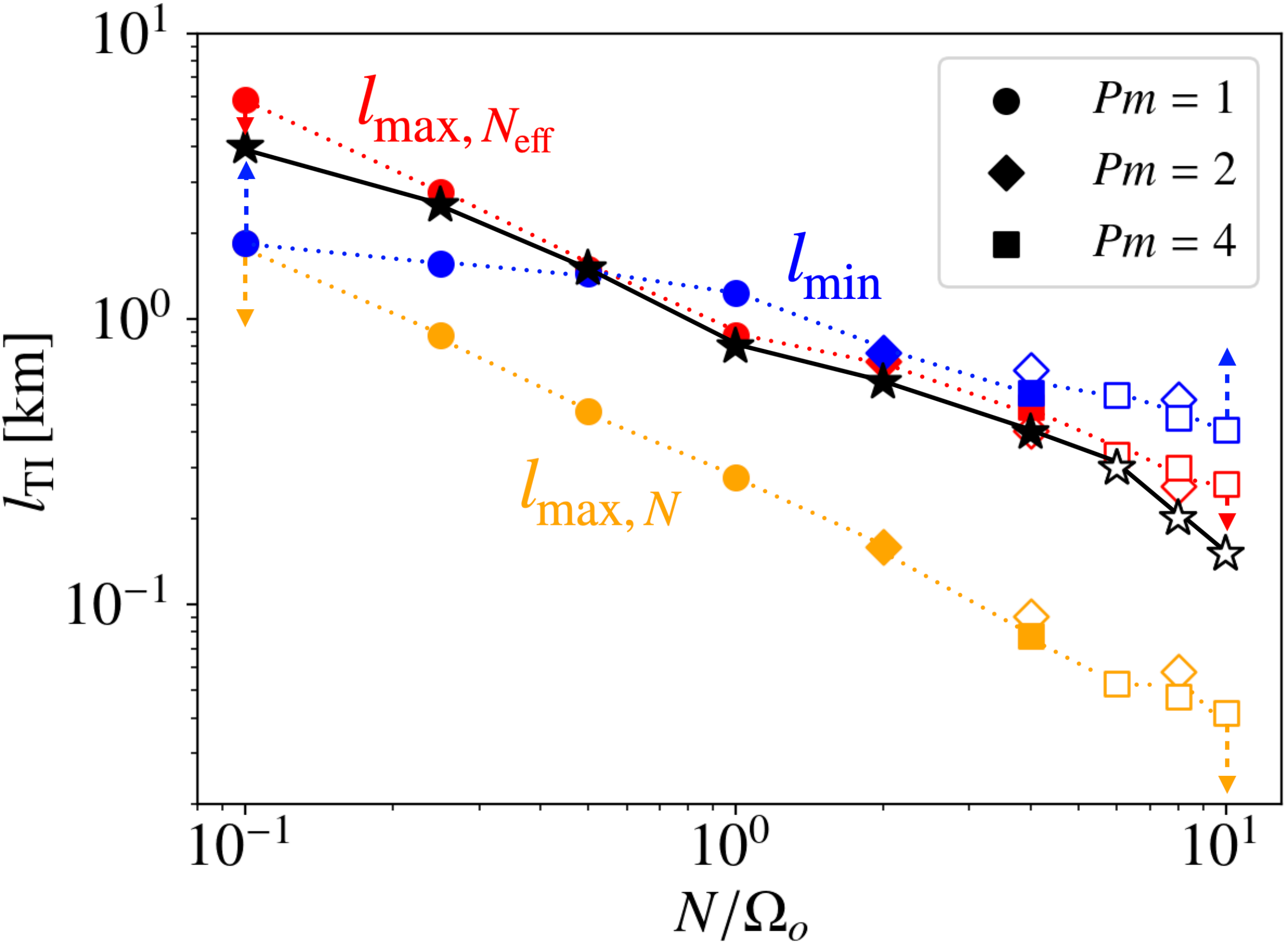}
    \caption{Length scale of the Tayler instability mode measured in the simulations (black stars) as a function of $N/\Omega_o$. The theoretical lower ($l_{\rm min}$ in blue) and upper boundaries of the length scale are also plotted using the classical ($l_{{\rm max},\,N}$ in orange) and the effective ($l_{{\rm max},\,N_{\rm eff}}$ in red) Brunt-V\"{a}is\"{a}l\"{a} frequencies. Filled and empty markers represent self-sustained and transient dynamos, respectively.}
    \label{fig:lTI}
\end{figure}

Moreover, the impact of stratification on the mode structure is striking. The stable stratification tends to stabilise displacements in the radial direction, as we can see looking at the non-axisymmetric radial velocity $v_r^{m\neq0}$ field in Fig.~\ref{fig:OmgVrp}. As a consequence, the radial length scale of the instability strongly decreases for increasing values of $N/\Omega_o$. This feature is not surprising because~\citet{spruit1999} already constrained the mode maximum radial length scale 
\begin{equation}\label{eq:lmax}
    l_{\rm TI}<l_{{\rm max},\,N}\equiv r\frac{\omega_{\rm A}}{N}\,,
\end{equation}
where $\omega_{\rm A}\equiv B_{\phi}^{m=0}/\sqrt{4\pi\rho r^2}$ is the Alfvén frequency. Note that a lower limit due to resistivity is also predicted
\begin{equation}\label{eq:lmin}
    l_{\rm TI}^2>l_{{\rm min}}^2\equiv\frac{\eta\Omega_o}{\omega_{\rm A}^2}\,.
\end{equation}
The length scales measured in our models are compared to these constraints in Fig.~\ref{fig:lTI}. Since thermal diffusion can mitigate the effect of stratification, we also define an effective Brunt-V\"{a}is\"{a}l\"{a} frequency 
\begin{equation}
    N_{\rm eff}\equiv N\sqrt{\eta/\kappa}
    = N \sqrt{Pr/Pm}
\end{equation}
and so
\begin{equation}\label{eq:lmax}
    l_{{\rm max},\,N_{\rm eff}}\equiv r\frac{\omega_{\rm A}}{N_{\rm eff}}\,
\end{equation}
to take this effect into account~\citep{spruit2002}. The Tayler modes in our simulations have length scales ranging from $r_o/4=\SI{3}{km}$ at $N/\Omega_o=0.1$ to $r_o/80=\SI{0.15}{km}$ at $N/\Omega_o=10$. This implies that the Tayler-Spruit dynamo requires higher and higher resolutions at greater stratifications to be resolved.  
The measured $l_{\rm TI}$ follows very well the upper limit $l_{{\rm max},\,N_{\rm eff}}$ (red points in Fig.~\ref{fig:lTI}), but is around one order of magnitude larger than $l_{{\rm max},\,N}$. This demonstrates the importance of including the mitigation of the stratification by diffusion. The minimum length scale $l_{\rm min}$ (Eq.~\eqref{eq:lmin}) is almost equal to $l_{\rm TI}$ from $N/\Omega_o=0.5$ to $N/\Omega=4$, which indicates that we are close to the instability threshold. For $N/\Omega_o\geqslant 6$, however, $l_{\rm min} \sim 2l_{{\rm max},\,N_{\rm eff}} \sim  2-3 l_{\rm TI}$. The fluid is therefore stable, which is consistent with the transient state  we find in our simulations. Thus, the analytical limits for the Tayler modes to develop are validated by our numerical simulations and suggest that the Tayler-Spruit dynamo could be maintained for $N/\Omega_o \in[6,10]$ with $Pm\gtrsim 16 - 36$.

\begin{figure}
    \centering
    \includegraphics[width=\columnwidth]{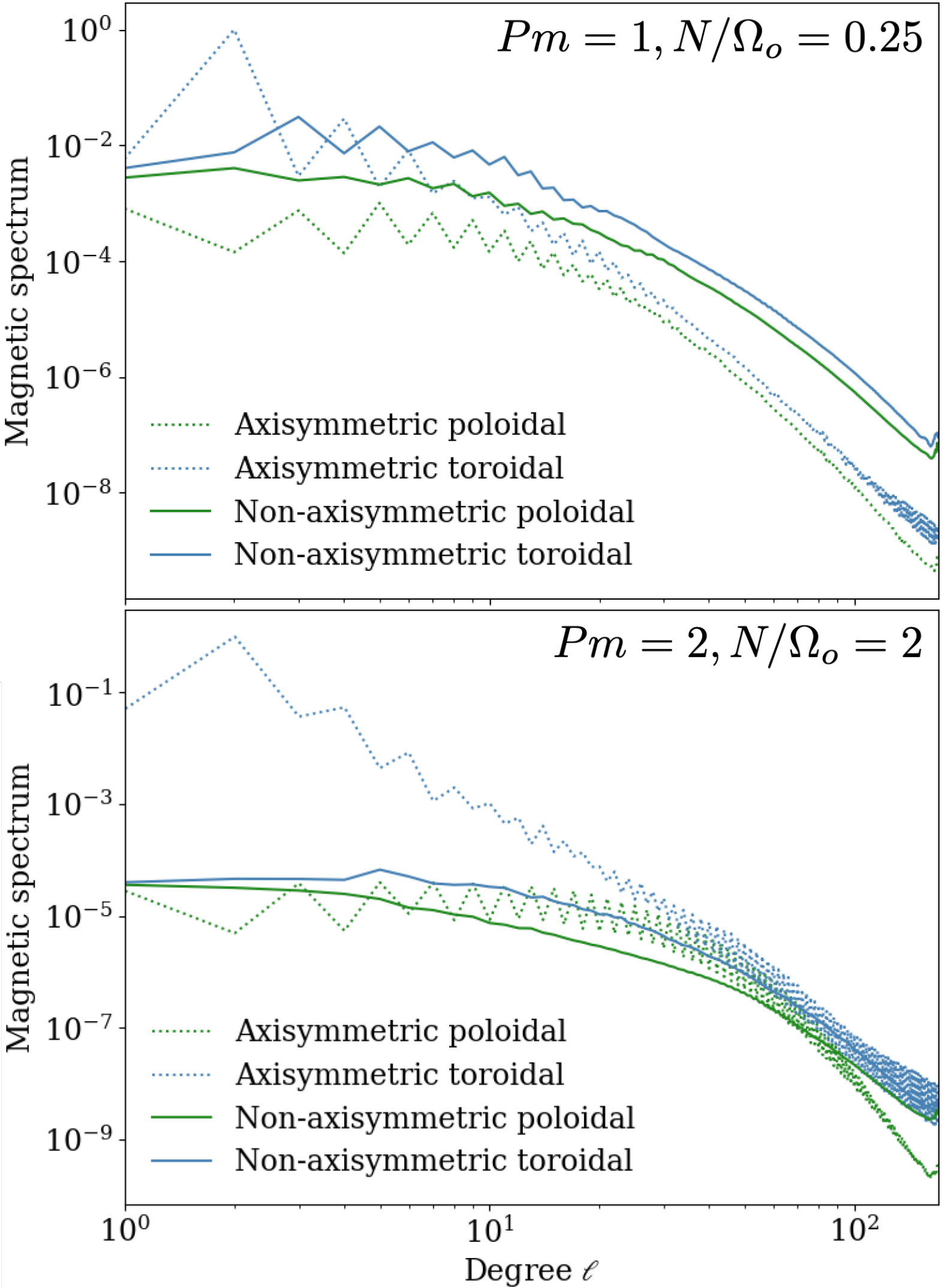}
    \caption{Time and volume averaged spectra of the magnetic energy for the parameters $Pm=1,N/\Omega_o=0.25$ (top) and $Pm=2,N/\Omega_o=2$ (bottom). The magnetic energy is normalized by  the energy of the dominant $(\ell=2,m=0)$-mode of the toroidal component.}
    \label{fig:spec}
\end{figure}

\begin{figure}
    \centering
    \includegraphics[width=\columnwidth]{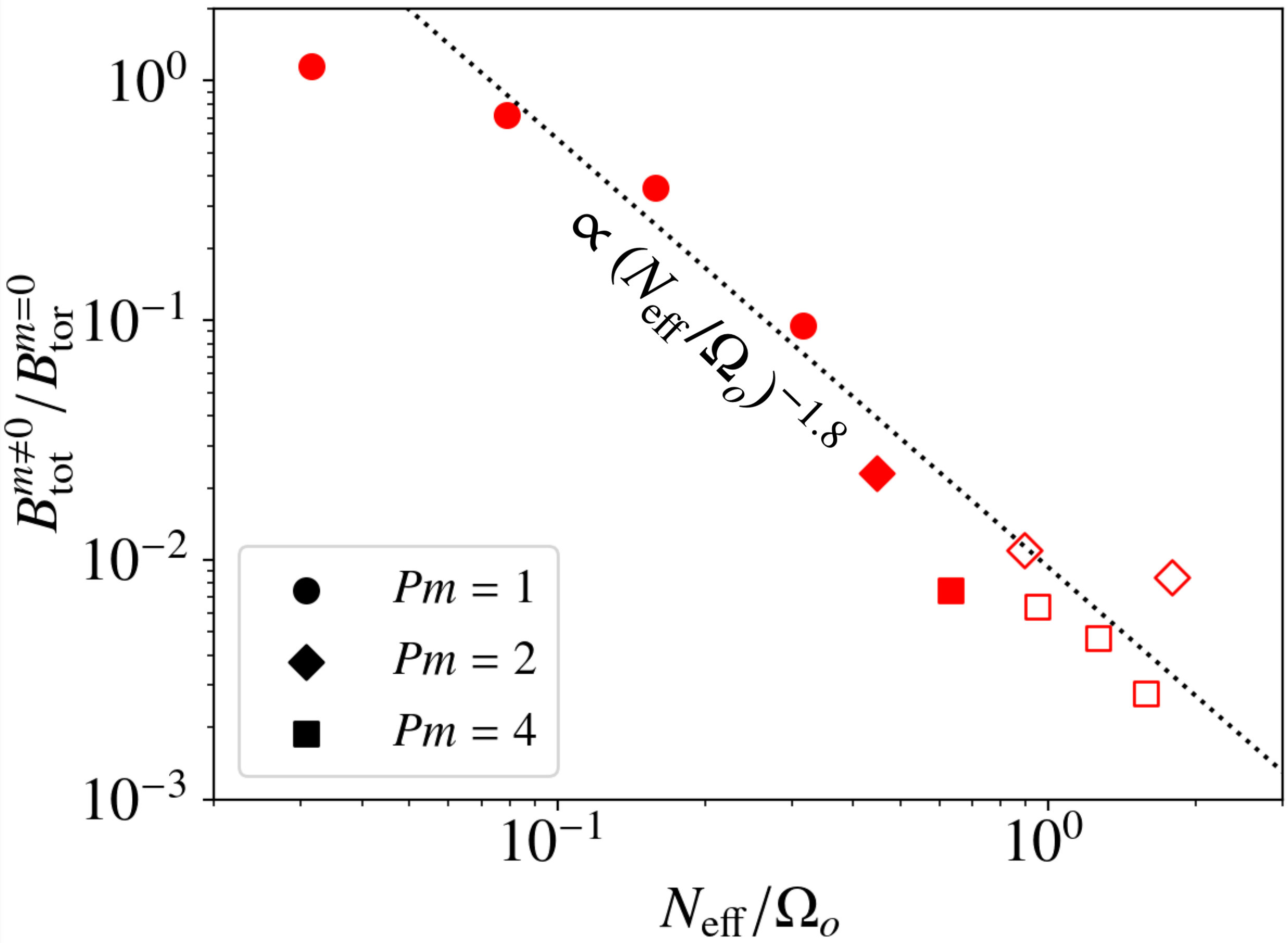}
    \caption{Ratio of the RMS non-axisymmetric magnetic field to the RMS axisymmetric toroidal magnetic field. The dotted line shows the best fit for a power law of $N_{\rm eff}/\Omega_o$. Filled and empty markers represent self-sustained and transient dynamos, respectively.}
    \label{fig:ratio_dBB}
\end{figure}

In addition to the decrease of $l_{\rm TI}$, the Tayler instability modes are strongly affected by high values of $N/\Omega_o$. The time and volume averaged spectrum of the magnetic energy in Fig.~\ref{fig:spec} show that the energy of the large-scale ($l=1-10$) non-axisymmetric modes (solid lines) drop by two orders of magnitude between $N/\Omega=0.25$ and at $N/\Omega=2$ compared to the energy of the dominant axisymmetric toroidal component (blue dotted line). This difference is represented more quantitatively by comparing the total non-axisymmetric magnetic field $B^{m\neq 0}_{\rm tot}$ to $B^{m=0}_{\phi}$ in Fig.~\ref{fig:ratio_dBB}. The ratio drops from $\sim 1$ to $\sim\SI{2e-3}{}$ and follows a power law $B^{m\neq 0}_{\rm tot}/B^{m=0}_{\phi}\propto N_{\rm eff}/\Omega_o^{-1.8\pm0.1}$. \citet{fuller2019}  analytically derived  that the ratio between the magnetic field generated by the Tayler instability (noted $\delta B_{\perp}\sim B^{m\neq 0}_{\rm tot}$) and $B^{m=0}_{\phi}$ follows $\omega_{\rm A}/\Omega_o$. Since $\omega_{\rm A}\propto N_{\rm eff}/\Omega_o^{-1/3}$~\citep[][and our Sect.~\ref{sec:scalings}]{fuller2019}, our simulations therefore do not match the analytical prediction. \citet{fuller2019} derived the ratio by equating the Tayler instability growth rate and a turbulent damping rate $\omega_{\rm A}^2/\Omega_o\sim\delta v_{\rm A}/r$, where $\delta v_{\rm A}\equiv \delta B_{\perp}/\sqrt{4\pi\rho}$. As the growth rate of the Tayler instability is robust~\citep{zahn2007,ma2019} and well verified in numerical simulations~\citep{ji2023}, our study then questions the prediction of the turbulent damping rate.

\subsection{Magnetic field saturation} \label{sec:scalings}
As in~\citet{barrere2023}, we confront the saturated large-scale magnetic fields in our simulations to the analytical predictions. To this end, we first measure the impact of the stratification on the local shear rate $q$, which influences the magnetic field saturation. Indeed, the rotation profiles of Fig.~\ref{fig:OmgVrp} show that the shear concentrates closer to the inner sphere and increases with $N/\Omega_o$. The quantification of this effect is described in Appendix~\ref{app:shear}. These larger values of $q$ explain the increase of the magnetic energy with $N/\Omega_o$ observed in Fig.~\ref{fig:bifurcation}.

In order to study the relation of the magnetic field components with $N_{\rm eff}/\Omega_o$ while taking into account the variation of $q$, we use the analytical prescriptions derived by~\citet{fuller2019}:
\begin{align}\label{eq:Bp_fuller}
    B^{m=0}_{\rm tor} & \sim \sqrt{4\pi\rho r_o^2}\Omega_o\left(\frac{q\Omega_o}{N_{\rm eff}}\right)^{1/3}, \\
    \label{eq:Br_fuller}
    B^{m=0}_{\rm pol},\, B_{\rm dip} & \sim \sqrt{4\pi\rho r_o^2}\Omega_o\left(\frac{q^2\Omega_o^5}{N_{\rm eff}^5}\right)^{1/3} \,.
\end{align}
The exponents of $q$ are all the more robust as they are confirmed by numerical simulations~\citep{barrere2023}. We define dimensionless magnetic field components compensated for the effect of the shear in the following way:
\begin{align}\label{eq:Bp_dimless}
    B^{m=0}_{\rm tor} & \longrightarrow \frac{B^{m=0}_{\rm tor}}{\sqrt{4\pi\rho r_o^2\Omega_o^2}q^{1/3}} \\ \label{eq:Br_dimless}
    B^{m=0}_{\rm pol},\, B_{\rm dip} & \longrightarrow \frac{B^{m=0}_{\rm pol}}{\sqrt{4\pi\rho r_o^2\Omega_o^2}q^{2/3}},\, \frac{B_{\rm dip}}{\sqrt{4\pi\rho r_o^2\Omega_o^2}q^{2/3}} \,.
\end{align}

\begin{figure}
    \centering
    \includegraphics[width=\columnwidth]{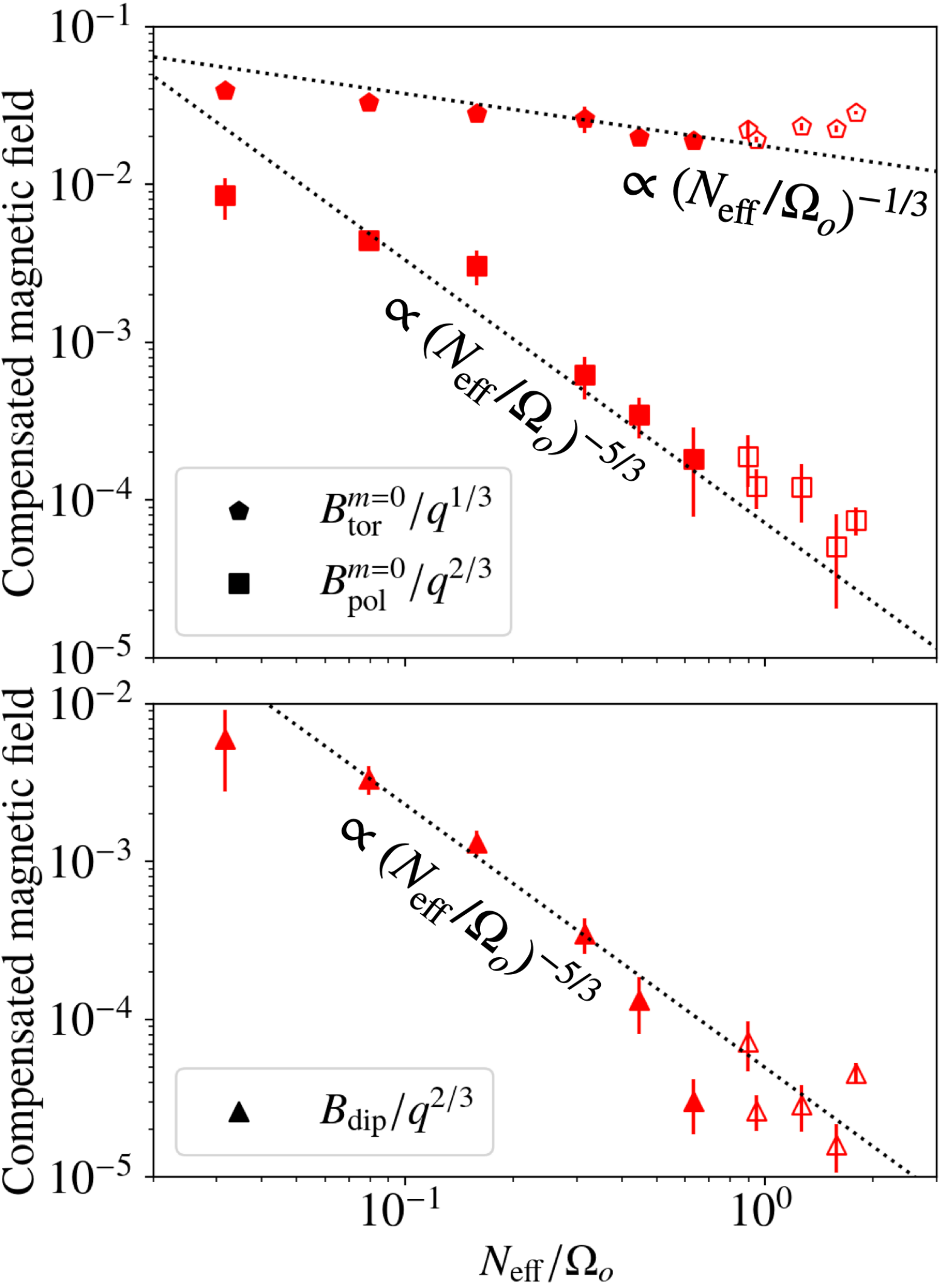}
    \caption{RMS toroidal and poloidal axisymmetric magnetic fields (top), and RMS magnetic dipole (bottom) compensated with the measured shear rate as a function of the ratio between the effective Brunt-V\"{a}is\"{a}l\"{a} frequency to the rotation rate at the outer sphere $N_{\rm eff}/\Omega_o$. The magnetic field is rendered dimensionless and compensated for the effect of the shear using Eqs.~\eqref{eq:Bp_dimless} and~\eqref{eq:Br_dimless}. Dotted lines show the best fits of the data with Fuller’s theoretical scaling laws (Eqs.~\eqref{eq:Bp_fuller} and ~\eqref{eq:Br_fuller}) within a multiplying factor. Filled and empty markers represent self-sustained and transient dynamos, respectively.}
    \label{fig:B_scalings}
\end{figure}

These compensated dimensionless components are plotted in Fig.~\ref{fig:B_scalings}as a function of $N_{\rm eff}/\Omega_o$. The theoretical scaling laws (dotted black lines) qualitatively match our data. Since the point at $N_{\rm eff}/\Omega_o=\SI{3e-2}{}$ diverges from the scalings due to the weaker effect of stable stratification, we exclude it while calculating the best fits.
We obtain the following power-laws $B_{\rm tor}^{m=0}\propto ({N_{\rm eff}/\Omega_o})^{-0.11\pm 0.05}$, $B^{m=0}_{\rm pol}\propto ({N_{\rm eff}/\Omega_o})^{-1.1\pm 0.2}$, and $B_{\rm dip}\propto ({N_{\rm eff}/\Omega_o})^{-1.5\pm 0.1}$. While $B_{\rm tor}^{m=0}$ and $B_{\rm pol}^{m=0}$ follow power-laws slightly less steep than predicted in Eqs.~\eqref{eq:Bp_fuller} and ~\eqref{eq:Br_fuller}, $B_{\rm dip}$ is in good agreement with Eq.~\eqref{eq:Br_fuller}.

\begin{figure}
    \centering
    \includegraphics[width=\columnwidth]{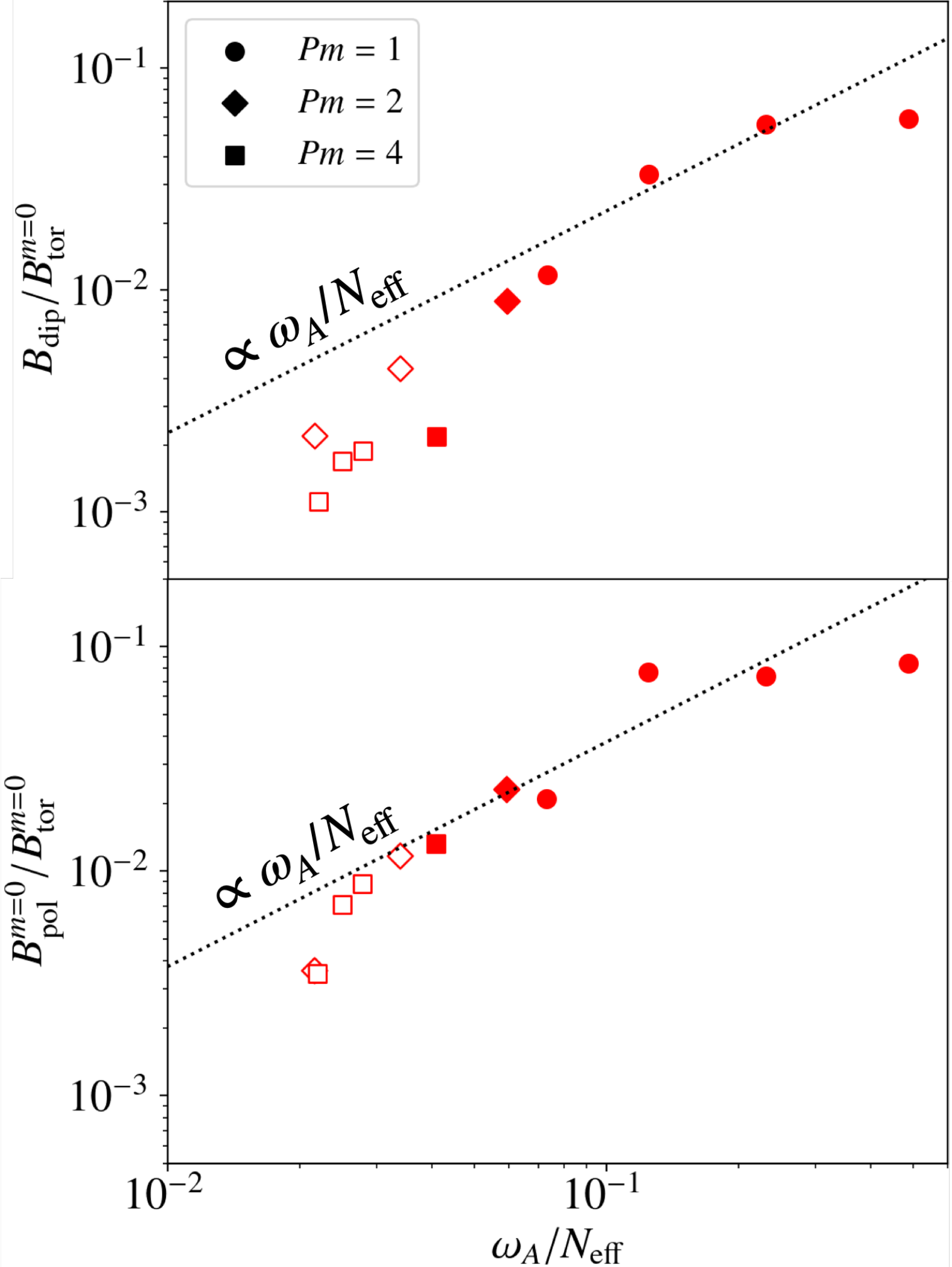}
    \caption{Ratio between the RMS axisymmetric poloidal (top) and the RMS dipolar (bottom) magnetic fields to the axisymmetric toroidal magnetic field. Dotted lines show the best fits of the data with Fuller’s theoretical scaling law $B_r/B_{\phi}\sim \omega_{\rm A}/N_{\rm eff}$ within a multiplying factor. Filled and empty markers represent self-sustained and transient dynamos, respectively.}
    \label{fig:ratio_scalings}
\end{figure}

This agreement with the theory is also found for the ratio $B^{m=0}_r/B^{m=0}_{\phi}\sim \omega_{\rm A}/N_{\rm eff}$~\citep{spruit2002,fuller2019} as seen in Fig.~\ref{fig:ratio_scalings}. Our data is fitted by the power law $B^{m=0}_{\rm pol}/B^{m=0}_{\rm tor}\propto (\omega_{\rm A}/N_{\rm eff})^{0.93\pm0.18}$, which is very close to the prediction. On the other hand, the ratio of the magnetic dipole to the axisymmetric toroidal field follows a somewhat steeper scaling law $B_{\rm dip}/B^{m=0}_{\rm tor}\propto (\omega_{\rm A}/N_{\rm eff})^{1.3\pm0.1}$.

\subsection{Angular momentum transport and mixing}\label{sec:AMT}
\begin{figure}
    \centering
    \includegraphics[width=\columnwidth]{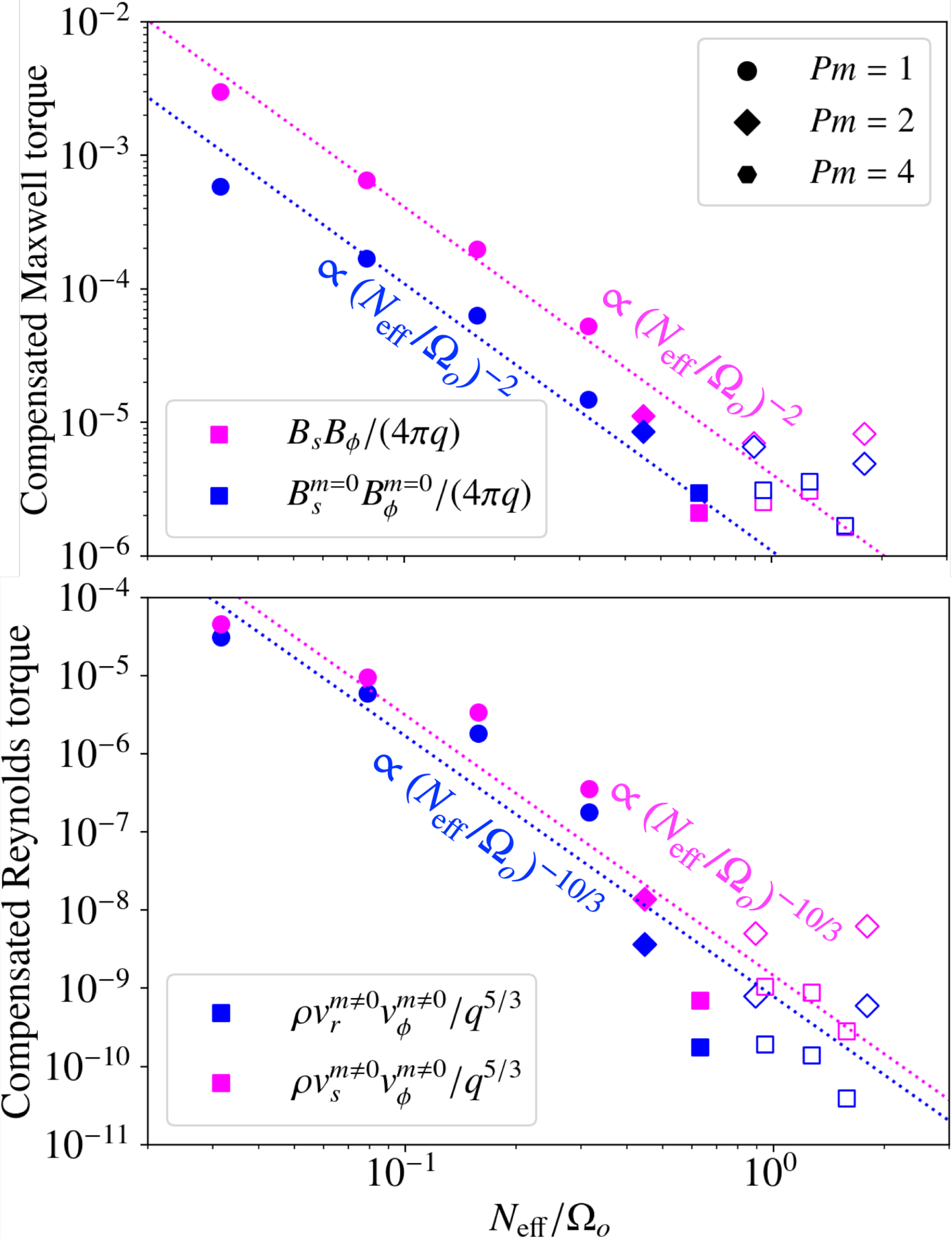}
    \caption{RMS Maxwell (top) and Reynolds (bottom) torques compensated with the measured shear rate as a function of the ratio between the effective Brunt-V\"{a}is\"{a}l\"{a} frequency to the rotation rate at the outer sphere. Dotted lines shows the best fits obtained with Fuller’s theoretical scaling laws. Filled and empty markers represent self-sustained and transient dynamos, respectively.}
    \label{fig:AMT_scalings}
\end{figure}
The angular momentum transport due to the large-scale magnetic field and turbulence in our simulations is also consistent with the theory of \citet{fuller2019}, as shown in Fig.~\ref{fig:AMT_scalings}. For the Maxwell torque $T_{\rm M}$, we find $B_sB_{\phi}\propto ({N_{\rm eff}/\Omega_o})^{-1.8\pm 0.1}$ and $B_s^{m=0}B_{\phi}^{m=0}\propto ({N_{\rm eff}/\Omega_o})^{-1.6\pm 0.1}$ depending on whether we take the non-axisymmetric components into account in $T_{\rm M}$. Note that the torque is more and more dominated by the axisymmetric magnetic fields as $N_{\rm eff}/\Omega_o$ increases. This dominance was assumed by~\citet{fuller2019} and can be expected given 
the results of Sect.~\ref{sec:TI}. The Reynolds torque values are more dispersed as a function of the stratification, but fit the power laws $v^{m\neq0}_rv^{m\neq0}_{\phi}\propto ({N_{\rm eff}/\Omega_o})^{-3.5\pm0.2}$ and $v^{m\neq0}_rv^{m\neq0}_{\phi}\propto ({N_{\rm eff}/\Omega_o})^{-3.4\pm0.2}$. Despite some scattering at high values of $N_{\rm eff}/\Omega_o$ in the points corresponding to transient dynamos, our data therefore follows well the analytical predictions $T_{\rm M}\propto ({N_{\rm eff}/\Omega_o})^{-2}$ and $T_{\rm R}\propto ({N_{\rm eff}/\Omega_o})^{-10/3}$ (dotted lines in Figs.~\ref{fig:AMT_scalings}). Moreover, we find $T_{\rm M}\sim\SI{e2}{}-\SI{e3}{}T_{\rm R}$, so the magnetic field is much more efficient than turbulence at transporting angular momentum. 

The mixing processes are also a crucial question in astrophysics, especially in stars. The Tayler-Spruit dynamo is expected to produce a very limited mixing efficiency compared to the angular momentum transport~\citep{spruit2002,fuller2019}. To measure this effect in our simulations, we define the effective angular momentum transport diffusivity $\nu_{\rm AM}\equiv T_{\rm M}/(\rho q \Omega_o)$ and roughly approximate the effective mixing diffusivity as $\nu_{\rm mix}\equiv q^{-5/3}v^{m\neq0}l_{\rm TI}$, with the rms turbulent velocity $v^{m\neq0}\equiv \sqrt{E_{\rm kin}^{m\neq 0}/(2\rho)}$ calculated from the mean non-axisymmetric energy $E_{\rm kin}^{m\neq 0}$. We divide by the power law $q^{5/3}$ in the expression of $\nu_{\rm mix}$ to take into account the variation of $q$ like in Figs~\ref{fig:B_scalings} and~\ref{fig:AMT_scalings}.

\begin{figure}
    \centering
    \includegraphics[width=\columnwidth]{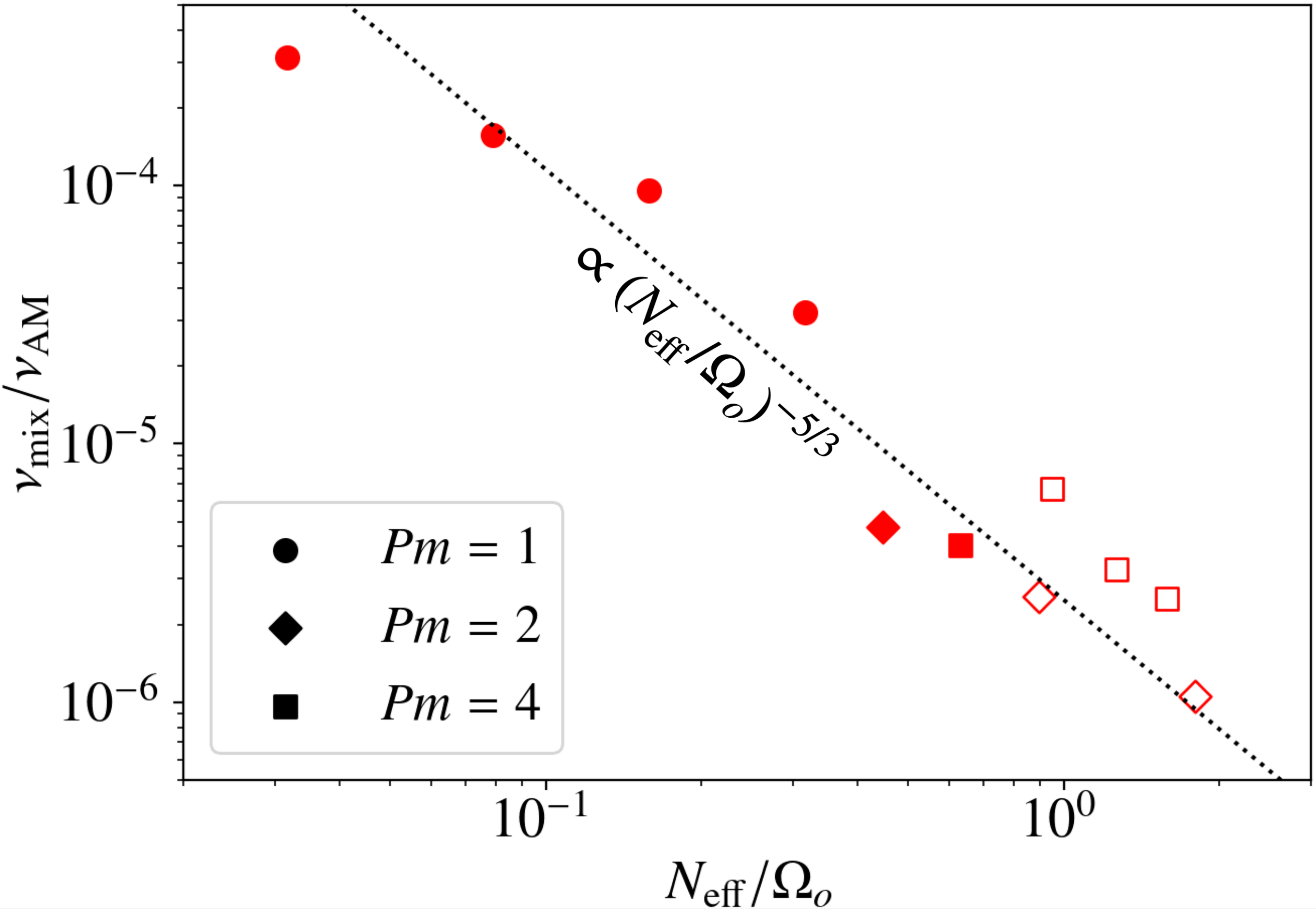}
    \caption{Ratio of the effective mixing diffusivity $\nu_{\rm mix}$ to the effective angular momentum diffusivity $\nu_{\rm AM}$ as a function of $N_{\rm eff}/\Omega_o$. Filled and empty markers represent self-sustained and transient dynamos, respectively.}
    \label{fig:nuMixAM}
\end{figure}

The ratio $\nu_{\rm mix}/\nu_{\rm AM}$ is plotted in Fig.\ref{fig:nuMixAM} and shows that our data is in fair agreement with the scaling $\nu_{\rm mix}/\nu_{\rm AM}\propto {N_{\rm eff}/\Omega_o}^{-5/3}$ of~\citet{fuller2019}. The power law $\nu_{\rm mix}/\nu_{\rm AM}\propto {N_{\rm eff}/\Omega_o}^{-1.2\pm0.2}$ best fits our data, which is mildly less steep than predicted. Moreover, our simulations also confirm that $\nu_{\rm mix}/\nu_{\rm AM}\sim \SI{e-6}{}-\SI{e-3}{}\ll 1$ for Tayler-Spruit dynamo. The use of passive scalars evolving in the velocity field in our simulations could help measure more precisely $\nu_{\rm mix}$ even though the approximation we used is satisfactory as a first analysis.

\begin{table*}
\centering
\caption{Table that sums up the theoretical and measured scaling laws of the different quantities discussed in Sects.~\ref{sec:scalings} and~\ref{sec:AMT}, and the dimensionless normalisation factor $\alpha$ defined by~\citet{fuller2019} (see Eq.~\ref{eq:alpha})}.
\begin{tabular}{lllr}
\hline
 Quantity (dimensionless) & \citet{fuller2019}'s scaling law & Best fit exponent & $\alpha$ \\
\hline
\hline
$ B^{m\neq0}_{\rm tot}/B^{m\neq0}_{\rm tor} $ & $ \omega_{\rm A}/\Omega_o $ & $ (N_{\rm eff}/\Omega_o)^{-0.18\pm0.1} $ & $ $ \\[0.2cm]
$ B^{m=0}_{\rm tor}/(\sqrt{4\pi\rho r_o^2\Omega_o^2}q^{1/3}) $ & $ \alpha (N_{\rm eff}/\Omega_o)^{-1/3} $ & $ (N_{\rm eff}/\Omega_o)^{-0.11\pm0.05} $ & $ 0.017 \pm 0.001 $ \\[0.2cm]
$ B^{m=0}_{\rm pol}/(\sqrt{4\pi\rho r_o^2\Omega_o^2}q^{2/3}) $ & $ \alpha^2 (N_{\rm eff}/\Omega_o)^{-5/3} $ & $ (N_{\rm eff}/\Omega_o)^{-1.1\pm0.2} $ & $ 0.009 \pm 0.002 $\\[0.2cm]
$ B_{\rm dip}/(\sqrt{4\pi\rho r_o^2\Omega_o^2}q^{2/3}) $ & $ \alpha^2(N_{\rm eff}/\Omega_o)^{-5/3} $ & $ (N_{\rm eff}/\Omega_o)^{-1.5\pm0.1} $ & $ 0.007 \pm 0.001 $ \\[0.2cm]
$ B^{m=0}_{\rm pol}/B^{m=0}_{\rm tor} $ & $ \omega_{\rm A}/N_{\rm eff} $ & $ (\omega_{\rm A}/N_{\rm eff})^{0.93\pm0.2} $ & --- \\[0.2cm]
$ B_{\rm dip}/B^{m=0}_{\rm tor} $ & $ \omega_{\rm A}/N_{\rm eff} $ & $ (\omega_{\rm A}/N_{\rm eff})^{1.3\pm0.1} $ & --- \\[0.2cm]
$ B_sB_{\phi}/(4\pi\rho r_o^2\Omega_o^2q) $ & $ \alpha^3 (N_{\rm eff}/\Omega_o)^{-2} $ & $ (N_{\rm eff}/\Omega_o)^{-1.8\pm0.1} $ & $ 0.016 \pm 0.004 $ \\[0.2cm]
$ B_s^{m=0}B_{\phi}^{m=0}/(4\pi\rho r_o^2\Omega_o^2q) $ & $ \alpha^3 (N_{\rm eff}/\Omega_o)^{-2} $ & $ (N_{\rm eff}/\Omega_o)^{-1.6\pm0.1} $ & $ 0.01 \pm 0.004 $ \\[0.2cm]
$ v_r^{m\neq 0}v_{\phi}^{m\neq 0}/(r_o^2\Omega_o^2q^{5/3}) $ & $ (N_{\rm eff}/\Omega_o)^{-10/3} $ & $ (N_{\rm eff}/\Omega_o)^{-3.5\pm0.2} $ & --- \\[0.2cm]
$ v_s^{m\neq 0}v_{\phi}^{m\neq 0}/(r_o^2\Omega_o^2q^{5/3}) $ & $ (N_{\rm eff}/\Omega_o)^{-10/3} $ & $ (N_{\rm eff}/\Omega_o)^{-3.4\pm0.2} $ & --- \\[0.2cm]
$ \nu_{\rm mix}/\nu_{\rm AM} $ & $ (N_{\rm eff}/\Omega_o)^{-5/3} $ & $ (N_{\rm eff}/\Omega_o)^{-1.2\pm0.2} $ & --- \\[0.2cm]
\hline
\end{tabular}
\label{tab:scaling_table}
\end{table*}

Table~\ref{tab:scaling_table} sums up the comparisons we have done between our data and the different scalings derived by~\citet{fuller2019}. Our results thus consolidate the validity of ~\citet{fuller2019}'s formalism
for the saturation of large-scale magnetic fields and angular momentum transport. Besides, our simulations are not compatible with the analytical prescriptions of~\citet{spruit2002}, which read
\begin{align}\label{eq:Bp_spruit}
    B^{m=0}_{\rm tor} & \sim \sqrt{4\pi\rho r_o^2}\Omega_o\left(\frac{q\Omega_o}{N_{\rm eff}}\right) \\
    \label{eq:Br_spruit}
    B^{m=0}_{\rm pol},\, B_{\rm dip} & \sim \sqrt{4\pi\rho r_o^2}\Omega_o\left(\frac{q^2\Omega_o^3}{N_{\rm eff}^3}\right) \\
    T_{\rm M} & \sim r_o^2\Omega_o^2q^3\left(\frac{\Omega_o}{N_{\rm eff}}\right)^4\,.
\end{align}

While our simulations support the scaling law of~\citet{fuller2019}, we can also constrain the dimensionless normalisation factor, (noted $\alpha$ in~\citet{fuller2019}), that parametrises the saturated strength of the axisymmetric toroidal magnetic field
\begin{equation}\label{eq:alpha}
    \frac{B^{m=0}_{\rm tor}}{\sqrt{4\pi\rho r_o^2}} = \alpha\Omega_o\left(\frac{q\Omega_o}{N_{\rm eff}}\right)^{1/3}\,.
\end{equation}
We infer the value of $\alpha$ by fitting our data by the theoretical scaling law. The measures are listed in the last  column of Table~\ref{tab:scaling_table} and we find a mean value of $\alpha\sim\SI{e-2}{}$. This value is small compared to those inferred by adjusting $\alpha$ in 1D stellar evolution models to the asteroseismic observations of sub-/red giants, which is $\sim0.25-1$~\citep{fuller2019,fuller2022,eggenberger2019b}. Either way, our numerical simulations provide a more physically motivated value of $\alpha$ that could be implemented in 1D stellar evolution codes including the Tayler-Spruit dynamo to transport angular momentum.

\subsection{Intermittency}\label{sec:intermittency}
\begin{figure}
    \centering
    \includegraphics[width=\columnwidth]{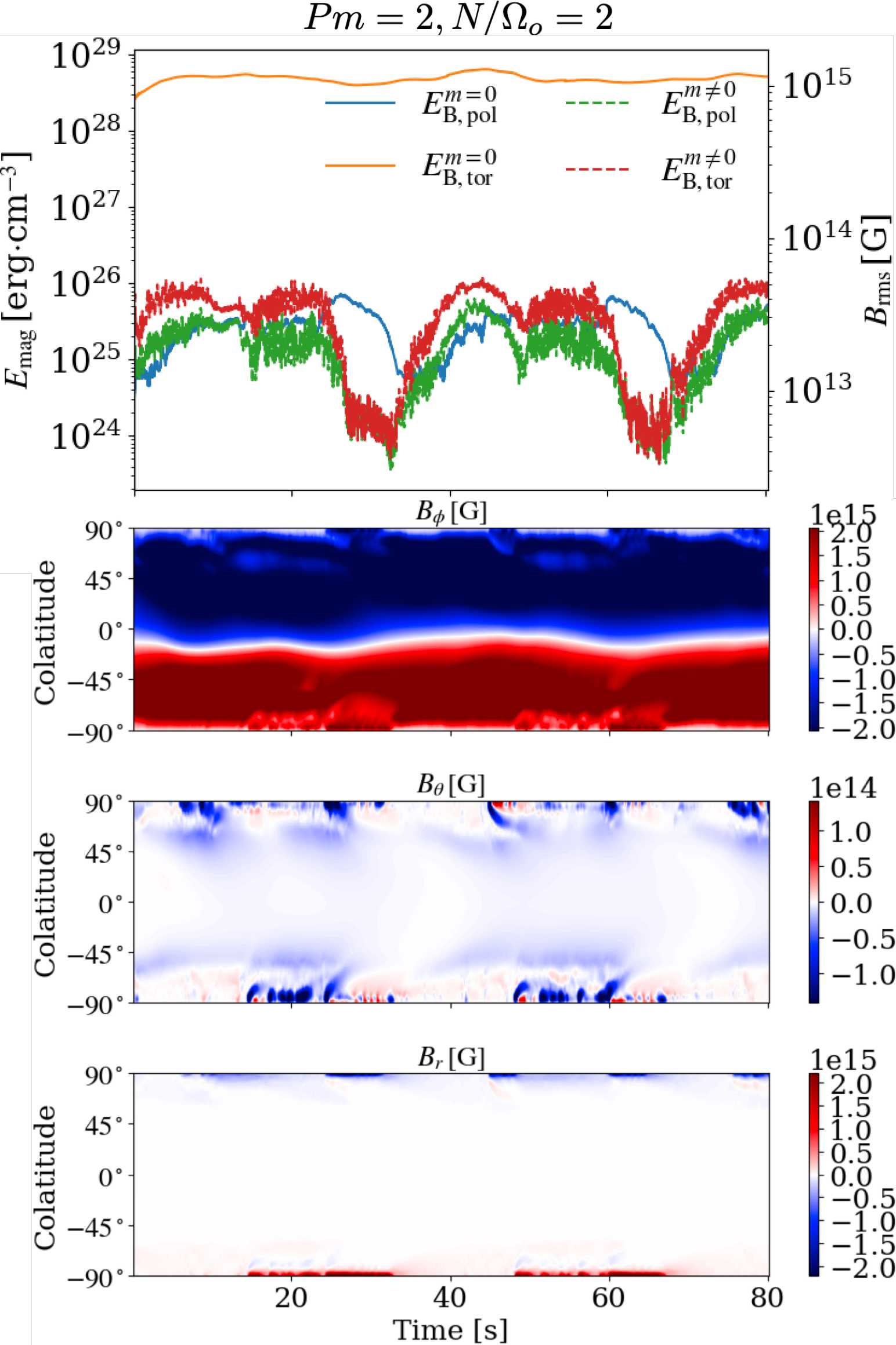}
    \caption{Top: Time series of the magnetic energy. Bottom: Butterfly diagram showing the latitudinal structure time evolution of different axisymmetric magnetic field components averaged between the radii $r=\SI{5}{}$ and $r=\SI{6}{km}$. The magnetic energy was converted to physical units by fixing $N=\SI{e-3}{s^{-1}}$}
    \label{fig:ts_buttB}
\end{figure}

When $N/\Omega_o\geq 2$, we find that the Tayler-Spruit dynamo displays an intermittent behaviour, which is clearly visible in the time series of Fig.~\ref{fig:ts_buttB} where the non-axisymmetric magnetic energy drops and increases cyclically by two orders of magnitude. This corresponds to the loss and growth of the Tayler instability. The same cycle also occurs for the axisymmetric $B_r$ and $B_{\theta}$, which illustrates the loss of the dynamo. Those two cycles show a very short lag of $\sim\SI{2.4}{s}$. We then notice that the oscillations of the axisymmetric toroidal and poloidal magnetic energies are in antiphase. This is also observed in the butterfly diagrams in which $B_{\phi}$ decreases locally, and so in the volume average when $B_r$ is the strongest. These cycles can be interpreted qualitatively as follows:
\begin{itemize}
    \item[(i)] $B^{m=0}_{\phi}$ is close but above the critical strength for the Tayler instability derived by combining Eqs.~\eqref{eq:lmax} and~\eqref{eq:lmin}
\begin{equation}
    B^{m=0}_{\phi,\rm c}\equiv\sqrt{4\pi\rho r_o^2}\Omega_o\left(\frac{N_{\rm eff}}{\Omega_o}\right)^{1/2}\left(\frac{\eta}{r_o^2\Omega_o}\right)^{1/4}
\end{equation}

and the dynamo is acting to generate $B^{m=0}_r$ ;
\item[(ii)] $B^{m=0}_{\phi}$ decreases slightly below the critical strength due to turbulent dissipation, which kills the Tayler instability and so the dynamo loop ;
\item[(iii)] the axisymmetric poloidal magnetic energy drops and the axisymmetric toroidal component increases because of the winding and the lack of turbulent dissipation ; 
\item[(iv)] $B^{m=0}_{\phi}$ exceeds the critical strength and the dynamo is active again. 
\end{itemize} 
An intermittent Tayler-Spruit dynamo was already proposed by~\citet{fuller2022} to explain the angular momentum transport in stellar stellar radiative regions with a low shear.
\begin{figure}
    \centering
    \includegraphics[width=\columnwidth]{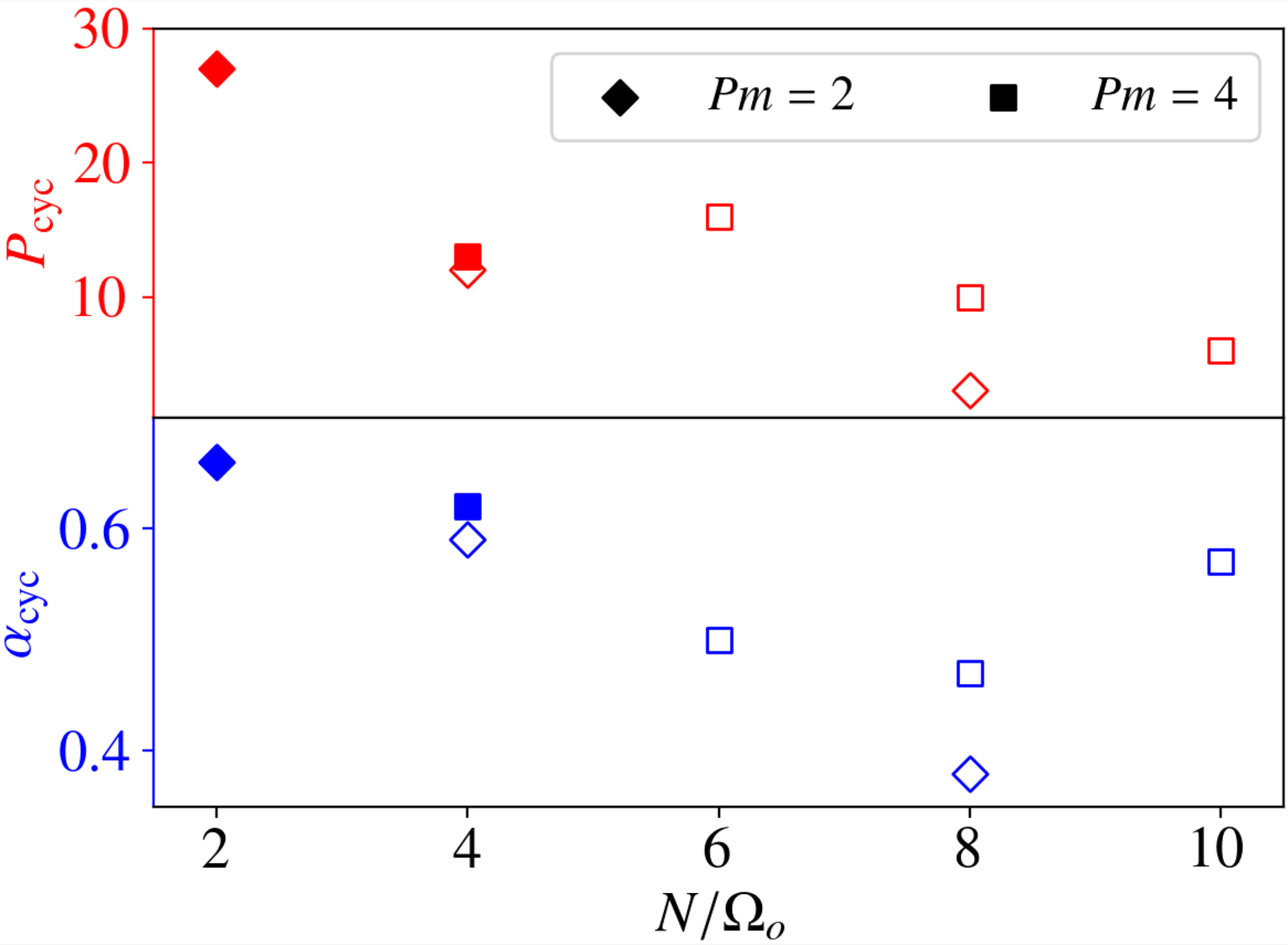}
    \caption{Period of the cycle $P_{\rm cyc}$ (top) and the duty cycle $\alpha_{\rm cyc}$ (bottom) of the intermittent dynamo as a function of the input $N/\Omega_o$. Filled and empty markers represent self-sustained and transient dynamos, respectively.}
    \label{fig:intermit_cyc}
\end{figure}

Quantitatively, we find $B_{\phi,\rm c}^{m=0}\sim\SI{1.4}{}-\SI{2.1e15}{G}$ for the models with $N/\Omega_o\in[2,10]$, which is very close to the maximum values $B_{\phi}^{m=0}\sim \SI{2.5}{}-\SI{3e15}{G}$ measured in the same models. The proximity to the instability threshold supports our interpretation. To characterise the time evolution of the intermittency, we measure its duty cycle $\alpha_{\rm cyc}$, i.e. the ratio of the time when the dynamo is active to the period of the cycle. We find that it varies between $0.38$ and $0.66$, with a tendency to decrease with $N/\Omega_o$ as seen in Fig.~\ref{fig:intermit_cyc}. The same trend is observed for the period of these cycles $P_{\rm cyc}$, which ranges between $\SI{3}{s}$ and $\SI{30}{s}$. This is consistent with the fact that we get closer to the dynamo threshold. 

\section{Application to magnetar formation}\label{sec:magnetars}

\begin{figure}
    \centering
    \includegraphics[width=\columnwidth]{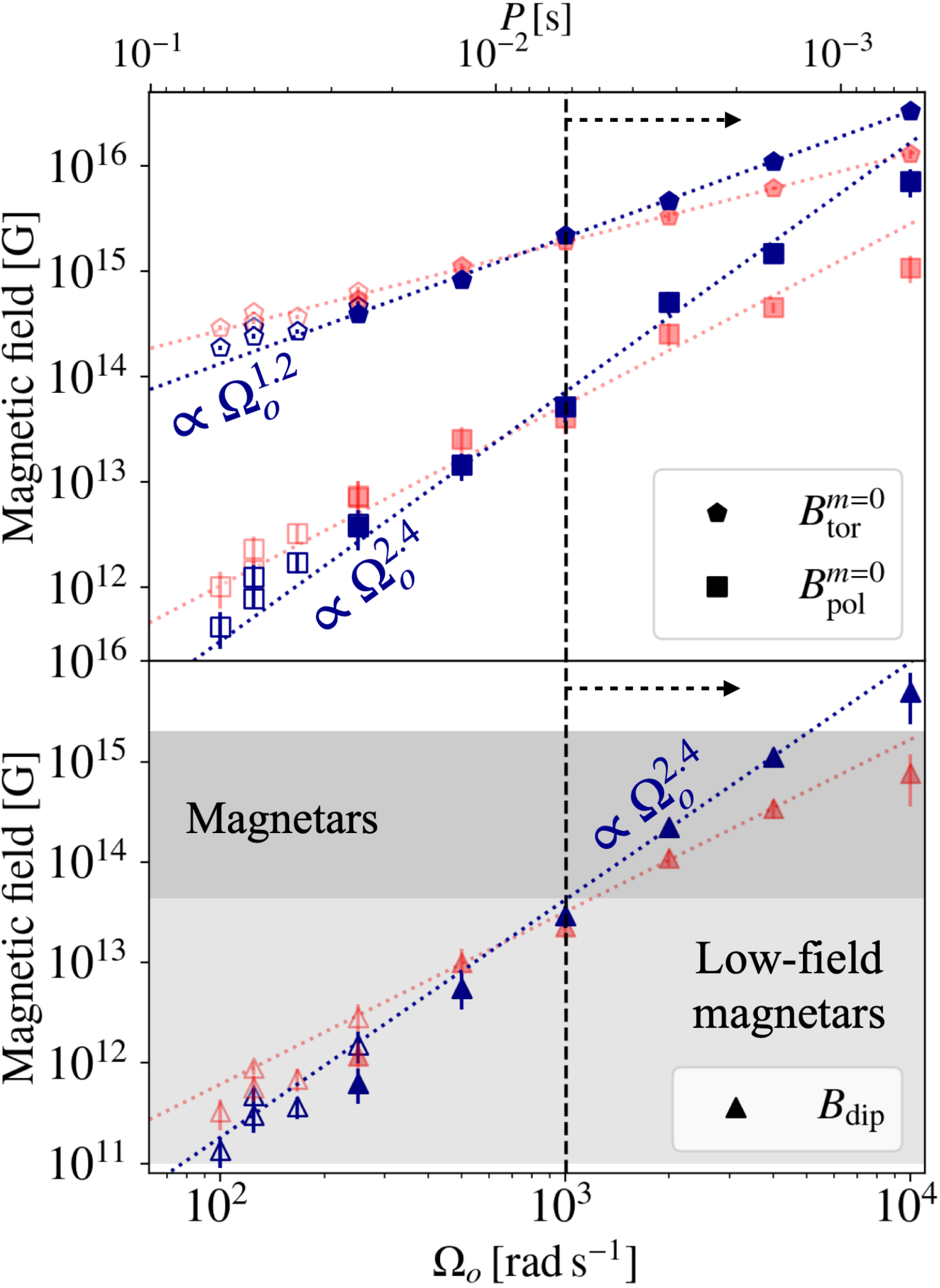}
    \caption{Magnetic strength of the axisymmetric toroidal $B^{m=0}_{\rm tor}$ (pentagons) and poloidal $B^{m=0}_{\rm pol}$ (squares) components (upper panel), as well as that of the magnetic dipole $B_{\rm dip}$ (triangles, lower panel) as a function of the angular frequency of the outer sphere, which represents the PNS surface. The red markers correspond to the magnetic field measured in the simulations, while the blue markers correspond to the values extrapolated to $q=1$. The dotted lines are the best power-law fit of the data. The dark and light grey regions represent the range of magnetic field for classical magnetars ($B_{\rm dip}\geqslant\SI{4.4e13}{G}$) and low-field magnetars ($B^{m=0}_{\rm tor}\geqslant\SI{e14}{G}$). The black dashed line and arrow illustrate the rotation period below which the dynamo can form classical magnetar-like magnetic fields. Filled and empty markers represent self-sustained and transient dynamos, respectively.}
    \label{fig:mag_app}
\end{figure}

In this section, we apply our numerical results to the magnetar formation scenario proposed by \cite{barrere2022}, whose semi-analytical modelling was based on the formalism by \cite{fuller2019}. To this end, the magnetic field is converted into physical units by fixing the following parameters to typical values in PNSs: the PNS radius $r_o=\SI{12}{km}$, mass $M=\SI{1.4}{M_{\odot}}$ that corresponds to a constant PNS density of $\rho\sim\SI{4.1e14}{g.cm^{-3}}$, and Brunt-V\"{a}is\"{a}l\"{a} frequency $N=\SI{1}{kHz}$. Fig.~\ref{fig:mag_app} shows the obtained magnetic field strength as a function of the angular frequency of the PNS surface, for the axisymmetric toroidal and poloidal components ($B^{m=0}_{\rm tor}$, $B^{m=0}_{\rm pol}$, upper panel) and for the dipolar component $B_{\rm dip}$ (lower panel). The red markers correspond to the magnetic field measured in the simulations, while the blue markers correspond to the values extrapolated to $q=1$ (as was assumed in \cite{barrere2022}). This plot can be compared to Fig.~5 in~\citet{barrere2022}, the main difference being that we define, here, a low-field magnetar as a magnetar with $B^{m=0}_{\rm tor}\geqslant\SI{e14}{G}$ but $B_{\rm dip}<\SI{4.4e13}{G}$. The magnetic field intensity show a similar trend with rotation frequency as was predicted by \cite{barrere2022}: the axisymmetric toroidal magnetic field increases as $\propto\Omega^{1.2}$ ($\propto\Omega^{4/3}$ in \cite{barrere2022}) while the poloidal and dipolar components increase as $\propto \Omega^{2.4}$ ($\propto\Omega^{8/3}$ in \cite{barrere2022}). This agreement is linked to the fact that the magnetic field in our numerical simulations follows well \citet{fuller2019}'s scaling law, as shown in the previous sections. The main difference is that the saturated magnetic field in our simulations is $\sim 17$ times weaker than in the model in~\citet{barrere2022}. This difference mainly comes from the assumption in \cite{barrere2022} of a dimensionless normalisation factor $\alpha = 1$, while our simulations indicate $\alpha\simeq0.01$ (see Table~\ref{tab:scaling_table}). Note that this value is smaller than the inverse of the measured dimensionless normalisation factor $\alpha^{-1}\sim 100$ because \citet{barrere2022} used $N$ instead of the $N_{\rm eff}$. 

The weaker magnetic field in our numerical simulations shifts the upper limit of the rotation period to form magnetar-like magnetic fields to $P\lesssim\SI{6}{ms}$. This new limit corresponds to a lower accreted fallback mass limit of $\sim\SI{5e-2}{M_{\odot}}$, which is still consistent with recent supernova simulations~\citep[see the discussion in][]{barrere2022}. For rotation periods longer than $\SI{6}{ms}$, the magnetic dipole is too weak for a classical magnetar but the Tayler-Spruit dynamo still produces strong total magnetic fields above $\SI{e14}{G}$. We suggest that these could correspond to the formation of low dipolar field magnetars. Indeed, the observations of absorption lines in the X-ray spectra of low-field magnetars~\citep{tiengo2013,rodriguez2016} and 3D numerical simulations of magnetic field long-term evolution in NSs~\citep{igoshev2021} suggest that such strength is enough to produce magnetar-like luminous activity.

\section{Discussion} \label{sec:discussion}
Here, we discuss the simplifications we used for the modelling of the PNS interior evolution: the mechanism to force the differential rotation (Sect.~\ref{sec:forcing}) and the Boussinesq approximation (Sect.~\ref{sec:boussinesq}). In Sect.~\ref{sec:comparison}, we finally compare our results on the Tayler-Spruit dynamo~\citep[][this article]{barrere2023} and the Tayler-Spruit dynamo obtained in other numerical simulations~\citep{petitdemange2023,daniel2023,petitdemange2024}.

\subsection{Dependence on diffusion processes}\label{sec:diff}

By assuming that all the rotational energy of the PNS is converted into kinetic energy of the explosions, observations of SN remnants associated to magnetars constrained a minimum initial PNS rotation period of $P_0\sim\SI{5}{ms}$~\citep{vink2006}. This value is close to our new constrain of the maximum $P$ to form magnetars through our scenario, which would leave only a small parameter space for the formation of magnetars by the Tayler-Spruit dynamo ($\SI{5}{ms}<P<\SI{6}{ms}$). However, we argue that our simulations may underestimate the strength of magnetic fields that would be generated in a realistic PNS. First, our simulations are far from the realistic regime of $Pm$, which reaches $Pm\sim\SI{e11}{}$ assuming a neutrino viscosity ($Pm\sim\SI{e4}{}$ assuming a shear viscosity) in PNSs $\sim\SI{10}{s}$ after the core bounce~\citep[see Supplementary Materials in][]{barrere2023}. Second, the thermal diffusivity is much larger than the resistivity in a PNS, which implies a small effective Brunt-V\"{a}is\"{a}l\"{a} frequency of $N_{\rm eff}=\SI{2.2e-8}{}N$ assuming a neutrino viscosity ($N_{\rm eff}=\SI{8e-4}{}N$ assuming a shear viscosity). The high-$Pm$ regime and the alleviation of the stable stratification favour most likely the development of the dynamo and stronger magnetic fields. The extrapolation of our results to the high Pm regime is an important open question, as it is unclear whether the theoretical scalings will hold in this regime.

\subsection{Forcing of the differential rotation}\label{sec:forcing}
To force the differential rotation, we chose to use a spherical Taylor-Couette configuration, in which a constant rotation rate is imposed on both inner and outer spheres. In this setup, the rotation profile is free to evolve as the angular momentum is transported by turbulence and large-scale magnetic fields. The imposed rotation of the outer sphere roughly mimics the maintenance of the surface rotation due to fallback accretion, once the PNS surface is already spun up significantly. However, the rotation profile evolution does not describe the beginning of the accretion during which the surface is spun up and the differential rotation, first concentrated close to the surface, is transported in the PNS interior. 

Maintaining the rotation on both spheres allows us to inject energy into the flow and try to control the shear rate. As noticed in Sect.~\ref{sec:scalings} and quantified in App.~\ref{app:shear}, the stable stratification however significantly changes the shear rate. 
This complicates the measure of the respective scaling exponents with $N/\Omega_o$ and $q$ independently. 
In addition, we observe in Fig.~\ref{fig:OmgVrp} that most of the shear is concentrated closer and closer to the inner sphere. As confirmed by our simulations, this restricts significantly the domain in which the Tayler-Spruit dynamo can operate and participate to make the dynamo more difficult to sustain. Thus, to investigate stronger stratification regimes, it will be necessary to change the forcing method and perhaps opt for a volumetric forcing as used for instance by~\citet{meduri2024}.

\subsection{Validity of the Boussinesq approximation}\label{sec:boussinesq}
To model the PNS interior, we used the Boussinesq approximation, which reduces the numerical cost and allows us to produce a few tens of models to better understand the physics of the Tayler-Spruit dynamo. Despite the importance of the density gradient, this approximation is reasonable in the case of PNS interior: 
\begin{itemize}
    \item[(i)] The sound speed is close to the speed of light $c_s\sim\SI{e10}{cm.s^{-1}}$~\citep[][private communication]{hudepohl2014,pascal2021}, so $v_{\rm A}/c_s \lesssim v_{\phi}/c_s \lesssim \SI{e-2}{}$, where $v_a \equiv r_o\omega_{\rm A}$ and $v_{\phi}$ are the typical Alfvén and azimuthal speeds.
    \item[(ii)] The density perturbation associated to the buoyancy term is small compared to the PNS mean density: $\delta\rho/\rho=\theta N^2/g\lesssim\SI{9e-2}{}$, with $N=\SI{e3}{s^{-1}}$, $g\sim G M/r_o\sim\SI{1.3e13}{cm.s^{-2}}$, and $\theta\lesssim r_o$ is the buoyancy variable (Eq.~\ref{eq:buoyancy}).
\end{itemize}
The impact of density gradient on the Tayler-Spruit dynamo has never been investigated so far in numerical simulations. Therefore, future work should consider more realistic PNS density profiles.

\subsection{Comparison with other numerical models}\label{sec:comparison}

In the literature, only a few other studies investigate numerically the Tayler-Spruit dynamo~\citep{petitdemange2023,petitdemange2024,daniel2023}. The main difference between our setup and theirs is the opposite shear, i.e. in their setup the inner boundary rotates faster than the outer one. As in our studies, they find a subcritical bifurcation at the Tayler instability threshold to a self-sustained state with a dominant axisymmetric toroidal magnetic field. However, many differences can be noticed:
\begin{itemize}
    \item The generated magnetic structure in their simulations has a smaller scale and is localized near the inner sphere in the equatorial plane. The impact of stable stratification on the length scale of these modes may deserve a deeper analysis. It is still unclear why this configuration is stable for one sign of the shear and not the other.
    \item As in~\citet{barrere2023}, a hemispherical dynamo solution is also found by~\citet{petitdemange2024} as they move from a laminar dynamo solution to the strong Tayler-Spruit dynamo by increasing $N/\Omega_o$. However, they do not find bistability between the hemispherical and the strong solutions as in~\citet{barrere2023}. 
    \item While the dipolar and hemispherical dynamos we found in~\citet{barrere2023} are in good agreement with the predictions of~\citet{fuller2019} and \citet{spruit2002}, respectively, all their models, including those of the hemispherical solution, are in agreement with the analytical model of~\citet{spruit2002}.
\end{itemize}
Therefore, the few numerical studies of the Tayler-Spruit dynamo indicate a much more complex physics than anticipated analytically, with the existence of a wide variety of dynamo solutions. So far, only~\citet{daniel2023} propose a non-linear model of the subcritical transition to the Tayler-Spruit dynamo of~\citet{petitdemange2023}. In order to include the other solutions we discovered, we must further investigate the dynamics of the dynamo using tools from dynamic system theory.

\section{Conclusions} \label{sec:conclusions}
\subsection{Summary}
Following our previous study~\citet{barrere2023} of the Tayler-Spruit dynamo with a fixed ratio of rotation to Brunt-V\"{a}is\"{a}l\"{a} frequency, we performed numerical simulations of the dipolar Tayler-Spruit dynamo to investigate its dependence on the level of stratification. The main results can be summarized as follows:
\begin{itemize}
\item We find self-sustained Tayler-Spruit dynamos for stratifications up to $N/\Omega_o = 4$ (and up to $N/\Omega_o= 10$ for transient dynamos). The dynamo also becomes intermittent as the saturated $B^{m=0}_{\phi}$ is close to the Tayler instability threshold for $N/\Omega_o\geqslant 2$. 
\item We observe a good agreement with the scaling laws of~\citet{fuller2019} for large-scale magnetic fields and the angular momentum transport, the latter of which is dominated by Maxwell torques.
\item With increasing $N/\Omega_o$, the Tayler modes have reduced radial length scales as expected but their energy decreases faster than theoretically predicted, which may indicate an underestimation of the turbulent dissipation by the analytical models. 
\item By measuring an approximate mixing diffusivity, we also determine the efficiency of the mixing process due to the Tayler-Spruit dynamo. We find that mixing is far less efficient than the angular momentum transport, as analytically predicted. 
\item Finally, as \citet{fuller2019}, we have defined a dimensionless normalisation factor $\alpha$ parameterising the scaling law of $B_{\rm tor}^{m=0}$ and numerically constrained its value, which is $\alpha\sim\SI{e-2}{}$. Therefore, the large-scale magnetic fields in our simulations are weaker than theoretically foreseen. 
\end{itemize}

Applying these results to the magnetar formation scenario of \cite{barrere2022}, the lower limit of the angular frequency to form classical magnetar-like dipoles is found to be larger than derived in~\citet{barrere2022} with a rotation period of $P\sim\SI{6}{ms}$. This value corresponds to an accreted fallback mass of $\sim\SI{5e-2}{M_{\odot}}$, which is still reasonable according to CCSN simulations~\citep[e.g.][]{sukhbold2016,sukhbold2018,chan2020,janka2021}. This new constraint is close to the minimum initial PNS period derived from the kinetic energy of SN remnants associated to magnetars~\citep{vink2006}. Nevertheless, as discussed in Sect.~\ref{sec:diff}, stronger magnetic fields are likely to form in the realistic PNS regime, alleviating this tension.

\subsection{Long-term evolution of the magnetic field}

After $\sim\SI{100}{s}$, the fallback accretion becomes too weak to maintain the differential rotation in the PNS. The newly formed strong large-scale magnetic fields transport the angular momentum efficiently, which damps the differential rotation and the dynamo will eventually stop. The magnetic field is expected to enter a relaxation phase in which its structure changes to reach a stable configuration. The exact shape of this magnetic field is still an open question and, more generally, the magnetic relaxation problem in astrophysics remains debated~\citep[e.g.][]{braithwaite2006,duez2010a,duez2010b,akgun2013,becerra2022b,becerra2022a}. It is however well acknowledged that the magnetic configuration is complex, mixing both large-scale poloidal and toroidal components. Thus, 3D numerical simulations including rotation and thermal/density stratifications are required to investigate this stage of the PNS evolution.

On longer timescales of $\sim 1-\SI{100}{kyr}$, the remaining strong toroidal magnetic fields located in the NS crust are prone to Hall diffusion and instability~\citep{rheinhardt2002}, which modifies their structures and so can influence the magnetar emission. The strong magnetic field-induced stresses could also cause failures or plastic deformations, which are suspected to explain the origin of magnetar bursts~\citep[e.g.][]{thompson1995,perna2011,lander2015,lander2019}.
It is therefore crucial to run 3D numerical simulations of magnetic field evolution in a NS structure using dynamo-generated initial magnetic configuration to better constrain these properties. Further investigations could also include the relaxation of the dynamo-generated magnetic field to a stable configuration before the PNS becomes a cooled stable NS.

\subsection{Interaction with a remaining fallback disc}
The magnetic dipole generated by the dipolar Tayler-Spruit dynamo may not be strong enough to spin the magnetar down to the observed $8-\SI{12}{s}$ via the magnetic spin-down mechanism. A good alternative would be the propeller mechanism~\citep[e.g.][]{gompertz2014,beniamini2019,lin2021,ronchi2022}. This operates when the magnetosphere is large enough to interact with the remaining fallback disc, i.e. when the Alfvén radius is larger than the corotation radius. In the propeller regime, the inner disk matter is accelerated to super-Keplerian velocity, which produces an outflow and so an angular momentum transfer from the magnetar to the disc. If this mechanism operates in some magnetars, the magnetic dipole which is inferred from the values of the NS rotation period and its associated derivative will be overestimated. It thus fosters numerical studies of the fallback matter in 3D simulations of core-collapse SNe and investigations on the evolution of the potential remaining disc. This will help constrain which progenitors are the best candidates to form magnetars via our fallback scenario.

\subsection{Implications for stellar physics}
Our findings are also of importance for the study of stellar radiative zones. Indeed, the scaling laws and the dimensionless normalisation factor $\alpha$ derived from our simulations could be implemented in 1D stellar evolution codes. Evolution models using the prescriptions of~\citet{fuller2019} have already been computed for sub-giant/red giant stars but with larger values of $\alpha\sim0.25-1$. These studies find a strong flattening of the rotation profile and conclude that the prescribed Tayler-Spruit dynamo can not explain both rotation profiles of sub-giant and red giant stars~\citep{eggenberger2019b}, which suggests that different angular momentum transport mechanisms occur during these two phases~\citep{eggenberger2019a}. The future asteroseismic measurements of the magnetic fields in stellar interiors with PLATO will be crucial to clarify the question of the transport mechanisms. Though the first measurements of magnetic fields in some red giant cores suggest a strong fossil field~\citep{li2022,li2023,deheuvels2023}, it will be essential to infer the asteroseismic signature of magnetic fields generated by the simulated Tayler-Spruit dynamos for the future observations. Evolution models including MHD instabilities effects were also performed in the case of massive stars to constrain the rotation rate of the remaining PNS or black hole~\citep{griffiths2022,fuller2022}. They suggest that the angular momentum transport by MHD instabilities is significant in every stage of the massive star evolution. This stresses the importance of performing 3D anelastic simulations with realistic background profiles of radiative zones at different evolution stages to better constrain the angular momentum transport and infer more robust rotation rates of stellar cores.
\begin{acknowledgements}
      This work was supported by the European Research Council (MagBURST grant 715368), and the  PNPS and PNHE programs of CNRS/INSU, co-funded by CEA and CNES. Numerical simulations have been carried out at the CINES on the Jean-Zay supercomputer and at the TGCC on the supercomputer IRENE-ROME (DARI project A130410317).
\end{acknowledgements}

\bibliographystyle{aa}
\bibliography{biblio}

%


\begin{appendix} 
\onecolumn{
\section{Measure of the shear rate}\label{app:shear}
The differential rotation is characterized by a dimensionless shear rate $q=r\partial_r {\rm ln}{\Omega}$. We define an effective shear rate based on the time average of the radial rotation profile in the saturated state at the colatitude of $\theta=\pi/8\,{\rm rad}$. We measure an average slope in the range of radii where half of the Tayler mode energy (approximated by the latitudinal magnetic energy $E_{B_{\theta}}$) is concentrated around its maximum. We chose this particular method because this range of radii is the region where the dynamo occurs. The measures are displayed in Fig.~\ref{fig:shear} (red plot) along with other measures made with different methods. Whatever the method used, we see that all the measures follow the same trend with an increase of $q\propto N$ until $N/\Omega_o=4$ after which the values of $q$ stay almost constant. 

\begin{figure}[h!]
    \centering
    \includegraphics[width=0.65\textwidth]{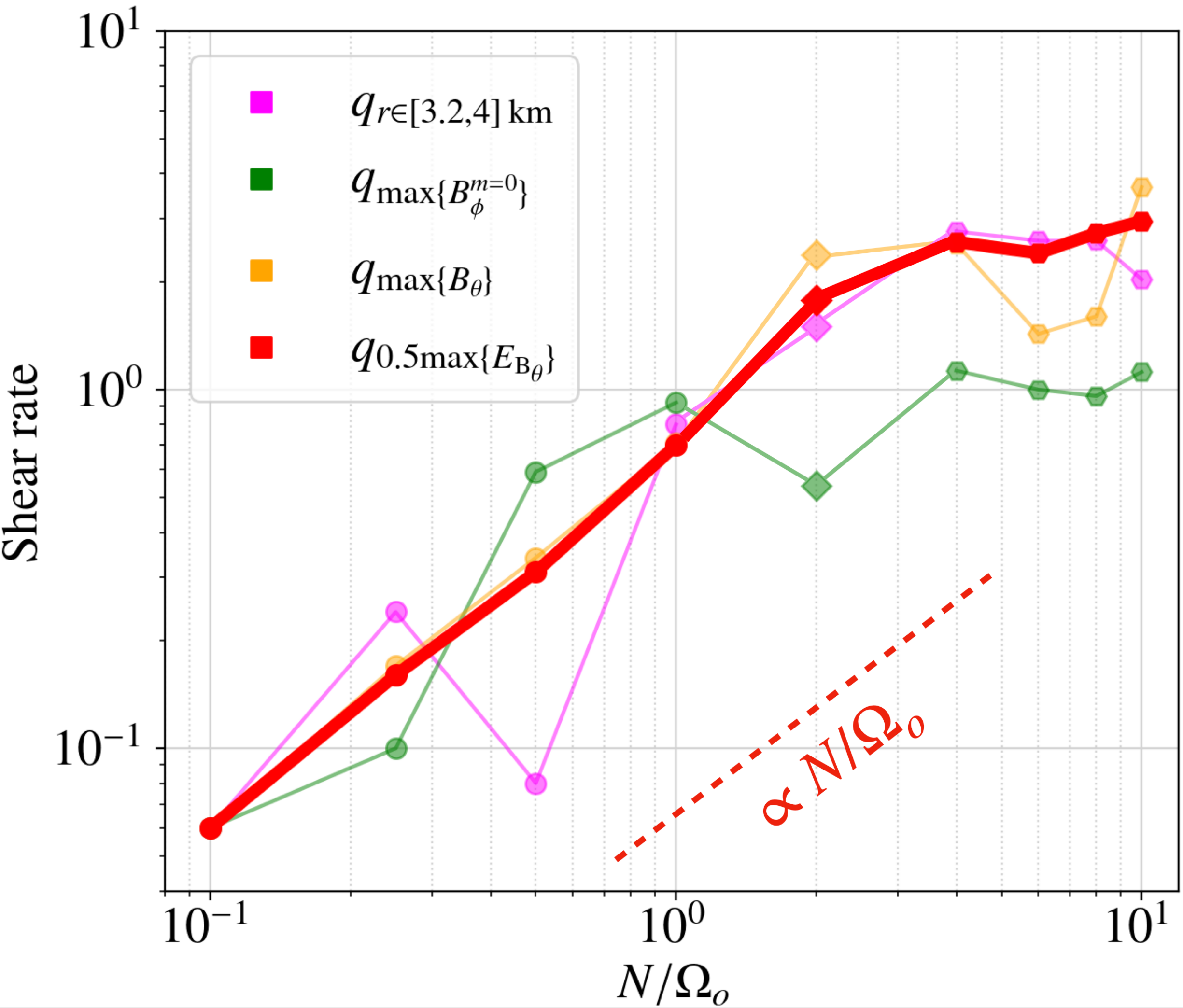}
    \caption{Shear rates $q$ measured locally in the simulations as a function of $N/\Omega_o$. The different colours represent distinct methods to measure $q$: slope in the rotation profile between $\SI{3.2}{}$ and $\SI{4}{km}$ (pink), $q$ at the maximum of $B^{m=0}_{\phi}$ and $B_{\theta}$ (green and orange, respectively), and slope in the range of radii where half of the Tayler mode energy (approximated by the latitudinal magnetic energy $E_{B_{\theta}}$) is concentrated around its maximum (red).}
    \label{fig:shear}
\end{figure}

\section{List of models}\label{app:models}
Tables~\ref{tab:sim_input}--\ref{tab:sim_output2} summarize the key parameters and output quantities of the simulations carried out in this study.

\begin{table*}[h!]
\caption{Overview of the stable (or failed) dynamo solutions. All the simulations have the same aspect ratio $\chi=0.25$, Ekman number $E=10^{-5}$, Rossby number $Ro=0.75$, thermal and magnetic Prandtl numbers $Pr=0.1$ and $Pm=1$, and the same resolution $(n_r,n_{\theta},n_{\phi})=(256,256,512)$.
Note that the run named $\text{Ro0.75s}$ is the same as in~\citet{barrere2023}. This table displays the input parameter of the runs.}
\begin{tabular}{lcccr}
\hline
 Name & $Pm$ & ${N/\Omega_o}$ & ${N_{\rm eff}/\Omega_o}$ & ${\Lambda_i}$ \\
   &   &   &   &   \\
\hline
$ \text{Ro0.75s} $ & $1$ & $0.1$ & $0.03$ & $ \text{$10$} $ \\
$ \text{Pm1Pr0.1NO0.25} $ & $1$ & $0.25$ & $0.08$ & $ \text{$\Lambda(\mathrm{Ro0.75s})$} $ \\
$ \text{Pm1Pr0.1NO0.5} $ & $1$ & $0.5$ & $0.16$ & $ \text{$\Lambda(\mathrm{Ro0.75s})$} $ \\
$ \text{Pm1Pr0.1NO1} $ & $1$ & $1$ & $0.32$ & $ \text{$\Lambda(\mathrm{Pm1Pr0.1NO0.5})$} $ \\
$ \text{Pm2Pr0.1NO2} $ & $2$ & $2$ & $0.45$ & $ \text{$\Lambda(\mathrm{Pm1Pr0.1NO1})$} $ \\
$ \text{Pm2Pr0.1NO4} $ & $2$ & $4$ & $0.89$ & $ \text{$\Lambda(\mathrm{Pm2Pr0.1NO2})$} $ \\
$ \text{Pm2Pr0.1NO8} $ & $2$ & $8$ & $1.79$ & $ \text{$\Lambda(\mathrm{Pm2Pr0.1NO4})$} $ \\
$ \text{Pm4Pr0.1NO4} $ & $4$ & $4$ & $0.63$ & $ \text{$\Lambda(\mathrm{Pm2Pr0.1NO4})$} $ \\
$ \text{Pm4Pr0.1NO6} $ & $4$ & $6$ & $0.95$ & $ \text{$\Lambda(\mathrm{Pm4Pr0.1NO4})$} $ \\
$ \text{Pm4Pr0.1NO8} $ & $4$ & $8$ & $1.26$ & $ \text{$\Lambda(\mathrm{Pm2Pr0.1NO8})$} $ \\
$ \text{Pm4Pr0.1NO10} $ & $4$ & $10$ & $1.58$ & $ \text{$\Lambda(\mathrm{Pm2Pr0.1NO10})$} $ \\
\hline
\end{tabular}

\label{tab:sim_input}
\end{table*}
\begin{table*}
\caption{Same as Table~\ref{tab:sim_input} but this table displays measured values in the simulations used to produce the plots of the paper.}
\begin{tabular}{lccccccr}
\hline
 Name & ${q}$ & $\Lambda$ & $B_{\mathrm{tor}}^{m=0}$ & $B_{\mathrm{pol}}^{m=0}$ & $B_{\mathrm{dip}}$ & $B_{\mathrm{tot}}^{m\neq 0}$ & $v_{\mathrm{tot}}^{m\neq 0}$ \\
   &   &   & $\left[10^{-3}\sqrt{4\pi\rho d^2\Omega_o}\right]$ & $\left[10^{-3}\sqrt{4\pi\rho d^2\Omega_o}\right]$ & $\left[10^{-3}\sqrt{4\pi\rho d^2\Omega_o}\right]$ & $\left[10^{-3}\sqrt{4\pi\rho d^2\Omega_o}\right]$ & $[10^{-4}d\Omega_o]$ \\
\hline
$ \text{Ro0.75s} $ & $0.06$ & $51.85$ & $21$ & $1.7$ & $1.2$ & $9.3$ & $35$ \\
$ \text{Pm1Pr0.1NO0.25} $ & $0.17$ & $69.03$ & $24$ & $1.8$ & $1.4$ & $9.7$ & $34$ \\
$ \text{Pm1Pr0.1NO0.5} $ & $0.35$ & $74.71$ & $26$ & $2.0$ & $0.87$ & $6.6$ & $36$ \\
$ \text{Pm1Pr0.1NO1} $ & $0.69$ & $99.44$ & $31$ & $0.64$ & $0.36$ & $2.6$ & $19$ \\
$ \text{Pm2Pr0.1NO2} $ & $2.37$ & $251.25$ & $36$ & $0.82$ & $0.31$ & $1.1$ & $6.0$ \\
$ \text{Pm2Pr0.1NO4} $ & $2.57$ & $332.08$ & $41$ & $0.49$ & $0.18$ & $0.61$ & $3.5$ \\
$ \text{Pm2Pr0.1NO8} $ & $2.54$ & $531.21$ & $52$ & $0.18$ & $0.10$ & $0.59$ & $3.4$ \\
$ \text{Pm4Pr0.1NO4} $ & $2.57$ & $478.89$ & $35$ & $0.46$ & $0.077$ & $0.36$ & $1.6$ \\
$ \text{Pm4Pr0.1NO6} $ & $2.59$ & $495.22$ & $35$ & $0.31$ & $0.077$ & $0.31$ & $1.8$ \\
$ \text{Pm4Pr0.1NO8} $ & $2.54$ & $714.16$ & $42$ & $0.31$ & $0.077$ & $0.28$ & $1.6$ \\
$ \text{Pm4Pr0.1NO10} $ & $3.74$ & $862.86$ & $47$ & $0.15$ & $0.051$ & $0.2$ & $1.1$ \\
\hline
\end{tabular}

\label{tab:sim_output1}
\end{table*}
\begin{table*}
\caption{Following of Table~\ref{tab:sim_output1}.}
\begin{tabular}{lccccccr}
\hline
 Name & $B_sB_{\phi}/4\pi$ & $B_s^{m=0}B_{\phi}^{m=0}/4\pi$ & $v_r^{m\neq 0}v_{\phi}^{m\neq 0}/4\pi$ & $v_s^{m\neq 0}v_{\phi}^{m\neq 0}/4\pi$ & $l_{\mathrm{TI}}$ & $P_{\mathrm{cyc}}$ & $\alpha_{\mathrm{cyc}}$ \\
   & $[10^{-6}\times 4\pi\rho d^2\Omega_o^2]$ & $[10^{-6}\times 4\pi\rho d^2\Omega_o^2]$ & $[10^{-8}\times\rho d^2\Omega_o^2]$ & $[10^{-8}\times\rho d^2\Omega_o^2]$ & [km] & [s$^{-1}$] &   \\
\hline
$ \text{Ro0.75s} $ & $15$ & $2.9$ & $52$ & $76$ & $2.8$ & -- & -- \\
$ \text{Pm1Pr0.1NO0.25} $ & $9.1$ & $2.4$ & $56$ & $89$ & $2.5$ & -- & -- \\
$ \text{Pm1Pr0.1NO0.5} $ & $5.7$ & $1.8$ & $58$ & $110$ & $1.5$ & -- & -- \\
$ \text{Pm1Pr0.1NO1} $ & $3.0$ & $0.84$ & $18$ & $35$ & $0.8$ & -- & -- \\
$ \text{Pm2Pr0.1NO2} $ & $2.2$ & $1.7$ & $2.8$ & $10$ & $0.6$ & $27$ & $0.66$ \\
$ \text{Pm2Pr0.1NO4} $ & $1.5$ & $1.4$ & $7$ & $4.4$ & $0.4$ & $12$ & $0.59$ \\
$ \text{Pm2Pr0.1NO8} $ & $1.7$ & $1.0$ & $0.52$ & $5.3$ & $0.2$ & $13$ & $0.38$ \\
$ \text{Pm4Pr0.1NO4} $ & $0.45$ & $0.63$ & $0.16$ & $0.61$ & $0.4$ & $3$ & $0.62$ \\
$ \text{Pm4Pr0.1NO6} $ & $0.53$ & $0.67$ & $0.17$ & $0.94$ & $0.3$ & $16$ & $0.5$ \\
$ \text{Pm4Pr0.1NO8} $ & $0.65$ & $0.75$ & $0.12$ & $0.77$ & $0.2$ & $10$ & $0.47$ \\
$ \text{Pm4Pr0.1NO10} $ & $0.50$ & $0.52$ & $0.066$ & $0.46$ & $0.15$ & $6$ & $0.57$ \\
\hline
\end{tabular}

\label{tab:sim_output2}
\end{table*}
}
\end{appendix}

\end{document}